\documentclass[10pt,twocolumn]{IEEEtran}
\usepackage{fullpage}
\usepackage{graphicx}
\usepackage[english]{babel}
\usepackage{multicol}
\usepackage{amsmath,amssymb}

\newcommand{\Et}[0]{E^{\text{Tx}}}
\newcommand{\Etl}[0]{E^{\text{Tx}}_{\ell}}
\newcommand{\EtS}[0]{E^{\text{Tx}}_{{S}}}
\newcommand{\Etot}[0]{E^{\text{total}}_{{S}}}
\newcommand{\EtR}[0]{E^{\text{Tx}}_{{R}}}
\newcommand{\avePER}[0]{\overline{\text{PER}}}
\newcommand{\MCS}[0]{{\text{MCS}}}
\newcommand{\dS}[0]{d_{S}}
\newcommand{\dR}[0]{d_{R}}
\newcommand{\dl}[0]{d_{\ell}}
\newcommand{\f}[1]{f\hspace*{-1mm}\left(#1\right)}

\begin{document}
\title{Energy-Throughput Trade-offs in a Wireless Sensor Network with Mobile Relay}
\author{
\authorblockN{Guanghua Zhu, Linda M.\ Davis, Terence Chan}

\authorblockA{Institute for Telecommunications Research\\
University of South Australia, Adelaide, Australia \\
\{Guanghua.Zhu, Linda.Davis, Terence.Chan\}@unisa.edu.au}

}

\maketitle

\begin{abstract}

In this paper we analyze the trade-offs between energy and throughput for links in a wireless sensor network.  Our application of interest is one in which a number of low-powered sensors need to wirelessly communicate their measurements to a communications sink, or destination node, for communication to a central processor.  We focus on one particular sensor source, and consider the case where the distance to the destination is beyond the peak power of the source.  A relay node is required.  Transmission energy of the sensor and the relay can be adjusted to minimize the total energy for a given throughput of the connection from sensor source to destination. We introduce a bounded random walk model for movement of the relay between the sensor and destination nodes, and characterize the total transmission energy and throughput performance using Markov steady state analysis. Based on the trade-offs between total energy and throughput we propose a new time-sharing protocol to exploit the movement of the relay to reduce the total energy. We demonstrate the effectiveness of time-sharing for minimizing the total energy consumption while achieving the throughput requirement.
% Link throughput, packet delay, and energy consumption per packet can be optimized using our time-sharing scheme.
%% Linda says: Little's law throughput = 1/average packet delay ... we actually do not optimize packet delay.  For any packet, the actual delay could be quite large.
We then show that the time-sharing scheme is more energy efficient than the popular sleep mode scheme.
\end{abstract}


\section{Introduction}

Wireless sensor networks (WSNs) have found wide application for delivering measurement observations in rural areas, battle fields, body area networks and other complex environments \cite{lewis2004wireless}. Research to date has been followed three main dimensions: sensing (e.g. sensor sampling \cite{Langtong}), processing (e.g. data aggregation \cite{Gobriel}, \cite{kalpakis2003efficient}), and communication (e.g. routing \cite{srinivas2003minimum} or data dissemination \cite{Selvakennedy}). Energy efficiency is a key design issue for all the three dimensions because most sensor nodes are battery powered. Batteries  often have small limited energy capacity and may be difficult to replace. When designing WSNs and the protocols, maximizing the network lifetime becomes one of the most significant problems \cite{goldsmith2002design}.

The objective in designing an energy efficient transmission protocol for WSNs is to deliver data from the sensor node to the destination efficiently using minimal energy, while still satisfying quality of service (QoS) constraints.
The choice of energy scheme depends on the application that the sensor network is intended to support. For real-time constant bit rate voice traffic or objective tracking scenarios, energy might be traded off to meet stricter QoS requirements \cite{so2009scheduling}, while for delay tolerant applications, such as rural area data acquisition systems (e.g. soil moisture sensing in \cite{shuman2010measurement}), protocols emphasizing network lifetime may be utilized.

One example of a wireless sensor network is the biological information acquisition system proposed in \cite{small2003shared}. The so-called \emph{shared wireless infestation model} (SWIM) was introduced to examine the whale data acquisition problems. Rather than the classic data collection model in which each whale transmits directly to a base station, the SWIM model permits sharing of information when whales are close to each other. Then, any of these whales that come into the coverage area of a base station can upload all the stored data, thus avoiding high transmission energy consumption at each whale node.
In a similar problem, \cite{estanjini2011optimizing} applied a WSN based system to a warehouse management model. Mobile forklifts (nodes) in the warehouse are fitted with sensor nodes which can collect information, including data related to the usage of the forklift, bumping/collision history, battery status, and  physical location. Information can be collected using IEEE 802.15.4 standard in an efficient event-driven manner and utilized to optimize forklift dispatching so as to minimize warehouse operational costs.

In \cite{grossglauser2002mobility}, Grossglauser and Tse proved that mobility of nodes in a network can greatly increase the network capacity or the per-user throughput. In fact,  two-hop routing (where data transverse to the sink node via at most one relay) was proved sufficient to attain the throughput capacity. It is worth noticing that in such a scenario, data will never be transmitted from a relay to another relay. Thus, cooperation among relays is not necessary. 

In this paper, our application of interest is one in which a number of low-powered sensors need to wirelessly communicate their measurements to a communications sink (a destination node or an \emph{infostation}~\cite{small2003shared}) labelled $D$.  This sink node is connected to some central processor for the measurement information via a backbone network, possibly wired. We will consider a specific (but not uncommon) scenario where the distance between the source sensor node $S$ and the infostation $D$ is so large that the power required by   $S$ to wirelessly communicate directly to $D$  is beyond the peak power of $S$.   Thus, a relay node  $R$, most likely another sensor node,   is necessary to assist the source to transmit data to the sink.  %This paper focuses on the trade-off between transmission throughput and delays of  one particular sensor source node $S$.

We introduce a bounded random walk model for movement of the relay node between the sensor and destination nodes. Following  a similar setup as in \cite{grossglauser2002mobility}, limited ``central coordinated scheduling'' is allowed in our model. In particular, we assume that the sink (i.e.\ the infostation) is not a typical sensor node. Hence, it is not power limited and can  send an almost instantaneous ACK to all sensor nodes to coordinate their transmissions. As we shall illustrate, this assumption can be relaxed in general. However, for analytical simplicity, it is assumed throughout the paper.  While we only focus on a single two-hop transmission from the sensor node $S$ to the infostation $D$, our model can also be extended to more general scenarios including the optimal two-hop routing scenario~\cite{grossglauser2002mobility}. Further details will be given in Section \ref{sec:ext}.

Most existing work (e.g., \cite{Urgaonkar:2011:NCR:2042997.2043013}\cite{Singh2010Delay}) on routing in delay tolerant networks use an ``on/off model'' for availability of an error-free link. As in \cite{Chen11:WInnFEu}, we note that the total energy consumption is related to both physical (PHY) layer parameters, including signal-to-noise ratio (SNR), and medium access control (MAC) layer parameters including packet length and transmission protocol parameters. One of the novelties in this paper is the incorporation of the error performance of the PHY layer in terms of the transmission energy in our link model.    This leads to a ÒsoftÓ characterization of the finite state machine, that is one in which transitions are present and the probability depends in a non-trivial way on the transmission energy and the hop distances between the source and relay, and the relay and destination. 
By understanding how the network throughput and transmission delays will be affected by the PHY and MAC  layers parameters, we derive design guidelines for how to choose or even optimize the PHY or MAC layers parameters in a delay tolerant networks.

Our work is also in contrast to the volume of research focused on the trade-offs in multiplexing and diversity for cooperative nodes in a relay network (usually with fixed transmission power). Instead, we focus on the energy-efficiency of communication and the trade-offs in energy consumption, throughput and relay mobility. Important questions include, what is the impact of the relay mobility on the link performance and what range of power levels are needed to meet the desired QoS parameters. One of the main novelties of our contribution is in formulation of the energy consumption using the Markov analysis to derive insights into non-trivial energy interactions in a simple relay link.   Transmission energy of the sensor node and the relay can be adjusted to minimise the total energy for a given throughput of the connection from sensor source to destination.

%
%In \cite{Cui04_JSAC,Cui05_TWC}, (this is about demonstrating the importance of circuitry energy)

Whether the relay is mobile or fixed, one common approach to energy saving in WSNs is to use sleep mode \cite{Ivan2009}.  Sleep mode exploits the fact that the sensor events are rare in some scenarios, and
 sensor nodes need to spend very little time in actual communication.
Sleep and wake up schemes typically trade off energy consumption for end-to-end delay and network connectivity.
A sleep mode based scheme called S-MAC was proposed in \cite{brownfield2006energy}, where nodes are organized into small groups called virtual clusters.
All the nodes in a virtual cluster have their sleep-wake schedule synchronized.
Multi-hop forwarding within the same virtual cluster, can take place without requiring to wait for the next-hop node to wake up.
The disadvantage of using such a scheme is that the co-ordination of the sleep-wake cycles requires some communication between the nodes, which in turn amounts to additional protocol and energy overheads.

An additional issue for WSN communication where retransmission is used is the size of the data packets.
Large packets reduce the probability of successful transmission while small packets lead to unnecessary overhead.  In  \cite{Chen11:WInnFEu}, a  variable packet length  adaptive modulation and coding  scheme was proposed to minimize the energy consumption for a transmit session over a direct wireless link.  Such a scheme is beyond the scope of this paper, but we note that it could be applied to the mobile relay system analyzed here.  In particular, we note that in the time-sharing scheme, packet lengths and modulation parameters could be adapted to suit the different power (i.e. SNRs) at the PHY layer.

Based on the trade-offs between total transmission energy and  throughput performance, we propose a time-sharing scheme with a high transmit energy mode and a low transmit energy mode. We show that such a scheme can exploit the movement of the relay to reduce the total energy consumption.  When packets are buffered for transmission, the time-sharing scheme can be more energy-efficient than a sleep mode scheme in meeting an average throughput or delay requirement.

The  paper is organized as follows: in Section \ref{sec:system}
%Sections \ref{sec:sys} and \ref{sec:nonthresh}, 
we present our detailed system model for the mobile wireless relay network, formulate the problem for optimization of energy consumption, and introduce the steady-state Markov analysis for calculating the two-hop throughput. In Section \ref{sec:ext} we discuss extensions and generalizations for the model.  Then, we present numerical results to characterize the throughput-energy trade-off in Section \ref{sec:char} for case study examples.  Based on observations from the optimisation in Section \ref{sec:opt}, we propose our time-sharing protocol and compare the time-sharing scheme with the popular sleep mode for energy saving in Section \ref{sec:MCTS}. We conclude the current work with our derived design guidelines in Section \ref{sec:application} and highlight some future directions in Section \ref{sec:concl}.

%
% We first consider a fixed relay position, putting in context our previous work \cite{zhu2011trade}.   We then present results for the mobile relay scenario.
%%%%%%%%%%%%%%%%%%%%%%%%%%%%%%%%%%

\section{Analytical model}\label{sec:system}

\subsection{Synchronization and Mobility}
\label{sec:sys}

In this section we present our system model and formulate the energy-throughput trade-off using a finite state model and Markov chain analysis.  We will focus on a particular sensor source $S$, and its connection to a communications sink or destination node $D$ via a relay node $R$  that acts as an intermediary between $S$ and $D$.  We assume that   $D$ is not within the transmission range of $S$, and hence  communications between $S$ and $D$ must be assisted by the relay node.  
%We model the  channels between the nodes as frequency-flat fading and constant during one packet data transmission. 
We assume that the transmission hops $S \rightarrow  R$ and $R  \rightarrow  D$ are independent and so do not interfere with each other. 
The destination node is a data collection infostation connected to a backbone network in order to communicate sensor measurement data from several sensor nodes to a central processor. It is  assumed to have sufficient power and hence can send an acknowledgement (ACK) to both $S$ and $R$ directly, whenever it successfully decodes and receives a packet.  Since we are interested in minimizing the energy consumed by the low-power devices $S$ and $R$ which is dominated by packet transmission (longer time scale), we assume that the ACK from $D$ (very short time scale) is effectively instantaneous.

In our model,  $S$, $R$, and $D$ do not need to be synchronized.  Nevertheless, we introduce the notion of a \emph{time slot interval} as a reference for the amount of energy used by the source and relay, as well as the amount of movement or relative speed of the mobile relay.
The communication protocol is divided into two phases. The first phase is for transmission from $S$ to $R$.  In this phase,  the sensor $S$ will transmit a packet of length $L$ channel input symbols to the relay $R$ using energy $\EtS L$.  The energy $\EtS$ describes the average energy per symbol used to transmit the $L$ symbol packet during the notional slot interval. It can represent a relatively high energy transmission of uncoded data or a lower energy transmission including redundancy coding.   The sensor $S$ will continue to transmit  the same packet until it receives an ACK from the destination node $D$.

The second phase begins when the relay $R$ successfully receives the packet sent by $S$.
In this second phase, the relay $R$ will transmit the packet to the destination node $D$. 
Note that in this phase, both $S$ and $R$ are transmitting, i.e.\ $S$ continues to transmit the packet even when the relay $R$ has already received it.  For the same packet of $L$ symbols, the relay uses energy $\EtR L$ in energy.   We assume the transmissions from $S$ and $R$ to be orthogonal, i.e.\  interference is non-existent or negligible.  This might be achieved through various multiple access schemes such as time-division, frequency-division, code-division multiple access or through directive antennas.  In fact, strict orthogonality is not required at all because $S$ and $D$ are so far apart that $D$ receives negligible signal from $S$, and sees only the signal from $R$.

Finally, when $D$ does decode the packet, an ACK will be broadcast to both $S$ and $R$.
%We assume the feedback channel for the ACK is error free and instantaneous (i.e.\  both S and R can receive the ACK/NACK immediately).
Upon receiving the ACK, the relay will stop forwarding the packet and $S$ will start a new transmission (entering the first phase again).

\begin{figure}[ht]
\centering
  \includegraphics[scale=0.85]{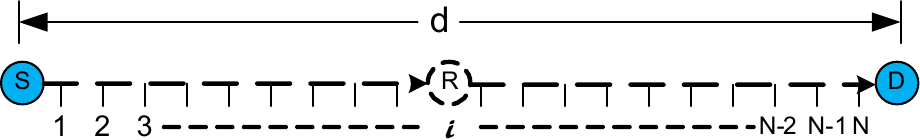}\\
  \caption{Link geometry for $S$ = sensor, $R$ = relay, and $D$ = destination nodes.  There are $N$ possible positions for the relay, current position, $i$.}\label{fig:systemodel}
\end{figure}

The distances and geometry between the nodes $S$, $R$ and $D$ affect the performance of the two hops in the link.  In general the three nodes form a triangle in a plane.  In the case when interference is negligible, it is only the distances that impact the model and so we can think of the nodes in a straight line for convenience, as shown in Figure \ref{fig:systemodel}.
Let $d$ be the distance between $S$ and $D$. We will  assume
that at any time instance, the relay must be at a distance of $\dS = id/(N+1)$ from the sensor for
an integer $i$ from 1 to $N$.  Similarly, the distance between $R$ and $D$ is given by $\dR = d - \dS = (1-i/(N+1))\,d$. We will simply say that  the relay is in  position $i$.

%\begin{figure}[ht]
%\centering
%  % Requires \usepackage{graphicx}
%  \includegraphics[scale=1.40]{brownianmotion.pdf}\\
%  \caption{Bounded Brownian motion of the relay node R with $s$ random walk steps taken between clock ticks $t$ and $t+1$.}\label{fig:brownian}
%\end{figure}

While both the sensor node $S$ and the destination node $D$ are fixed, the relay $R$ is mobile. In particular, we assume that the motion of $R$ follows  a bounded random  walk  model \cite{Childers97}.
Let $P $ be a $N\times N$ matrix such that its entry $P[i,j]$ at the $i^{th}$ row and the $j^{th}$ column of $P$ is defined as the conditional probability that $R$ is in position $j$ at the next step given that $R$ is in position $i$ at the current step.
% \begin{equation}
%P[i,j] \triangleq \Pr\left( 
%\begin{array}{l} R \text{ is in position }j \text{ at the next step } \\ | R \text{  is in position }i\text{ at the current step}
%\end{array}
%\right).
%%\label{eq:pij}
%\end{equation}
We assume the random walk is ``symmetric'' such that 
\begin{align}\label{eq:1}
P[i,j]
=\begin{cases}
1- p_{\rm move} & \text{ if } j=i \\
p_{\rm move}/2 & \text{ if } |j-i|=1, \: 1<i<N \\
p_{\rm move} & \text{ if } (j=2 \text{ and }  i=1)  \\ 
& \quad \text{ or }(j=N-1 \text{ and }  i=N) \\
0 & \text{ otherwise. }
\end{cases}
\end{align}

Roughly speaking, $p_{\rm move}$ is the probability that the relay $R$ will move in the slot period. Within each time slot (the time in which the source uses $\EtS L$ to transmit a packet, and the relay if transmitting, uses $\EtR L$), the relay will move $s$ steps. 
Hence, $s$ is essentially the mobility or speed parameter: the larger the value of $s$, the faster the relay is moving.  Let $P^{s}$ be the $s^{th}$ power of $P$ whose 
entry at the $i^{th}$ row and the $j^{th}$ column
is denoted as $P^{s}[i,j] $. Then $P^{s}[i,j] $ is the conditional probability that $R$ is in position $j$ after $s$ steps given that $R$ is in position $i$ at the current step.

%
%\begin{equation}
%P^{s}[i,j] \triangleq \Pr( R \text{ is in position }j \text{ after $s$ steps }| R \text{  is in position }i\text{ at the current step}).
%%\label{eq:pij}
%\end{equation}
% 

%when $1<i<N$. On the other hand, if $i=1 $ or $N$, then
%\begin{align*}
%p_{i,j}
%=\begin{cases}
%1- p_{\rm move} & \text{ if } j=i \\
%p_{\rm move} & \text{ if } (j=2 \text{ and }  i=1) \text{ or }(j=N-1 \text{ and }  i=N).
%\end{cases}
%\end{align*}

As we shall demonstrate, relay mobility  can significantly reduce the energy required to transmit a packet (especially in the low throughput regime). To see the idea, consider the scenario where both the sensor node $S$ and the relay node $R$ have a very limited transmission range (and hence also consume extremely low power).
Transmission from the  sensor to the relay is successful only when the relay node moves very close to the sensor node.
Even after receiving a packet from $S$, the relay node $R$ cannot successfully forward the packet to the destination node $D$ due to the long distance from $R$ to $D$. Instead, it wanders between $S$ and $D$. Clearly, it will get very close to the destination by chance after a period of time.
Then the packet being forwarded by $R$  can successfully be received at the destination. This approach may require a long time for a packet to be sent from $S$ to $R$ and then from $R$ to $D$. Despite its low throughput, our numerical results however show a huge reduction in packet transmission energy, that is, the total energy used by the source $S$ to successfully transmit the packet between $S$ and $D$.

Now, recall that $\EtS$  is the \emph{average symbol energy}, that is,  the average amount of energy required to transmit one channel input symbol. The level of $\EtS$ is selected within the range of $ 0 \le \EtS \le E_{max}$, where $ E_{max} $ depends on battery  peak power. Let $\tau$ be the \emph{packet transmission delay} which is defined as  the expected number of time slots needed to successfully transmit a packet between $S$ and $D$. Then,  $\Etot = \EtS \times L\times \tau  $ will be the expected total packet transmission energy needed by the sensor to successfully transmit a packet.

 In this paper, we focus on the amount of transmission energy consumed by the sensor source in calculating a total energy cost.  The model can be modified and extended to include the energy of the relay node used for transmission, even mobility, as well as the energies consumed by the relay node $R$ and the destination node $D$ to receive a packet.  We focus only on $S$ to simplify the exposition.  We will discuss this further in Section \ref{sec:ext}.

%\subsection{Markov Chain Formulation}
%\label{sec:MC}
To study the trade-offs among packet transmission energy, mobility and system throughput, we will use a Markov chain to model the whole communications process. The chain has $3N$ states which are labelled by
\begin{equation}
\Omega \triangleq \{\mathbb F^{\,i}, \mathbb R^{\,i}, \mathbb B^{\,i} , i =1 , \ldots, N \}.
\end{equation}
The state in which the communications system is in depends on 1) which nodes ($S$ only, or both $S$ and $R$) are transmitting, 2) whether it is a new packet transmission or not, and 3) the position of the relay node. The precise definition of the states are given as follows.

 \begin{itemize}

 \item  State $\mathbb F^{\,i}$: The system is in state $\mathbb F^{\,i}$ if (1) the relay $R$ is in position $i$, and (2) the sensor  $S$ is transmitting a new packet for the FIRST time.

 \item State $\mathbb R^{\,i}$: The system is in State $\mathbb R^{\,i}$ if (1) the relay is in position $i$, and (2) only the sensor is RETRANSMITTING the packet (because the relay  fails to decode the packet previously sent by the sensor S).

\item State $\mathbb B^{\,i}$:  The system is in State $\mathbb B^{\,i}$ if (1) the relay is in position $i$, (2) BOTH  $S$ and $R$ are transmitting the packet.

\end{itemize}

\begin{figure}[ht]
\centering
  \includegraphics[scale=0.4]{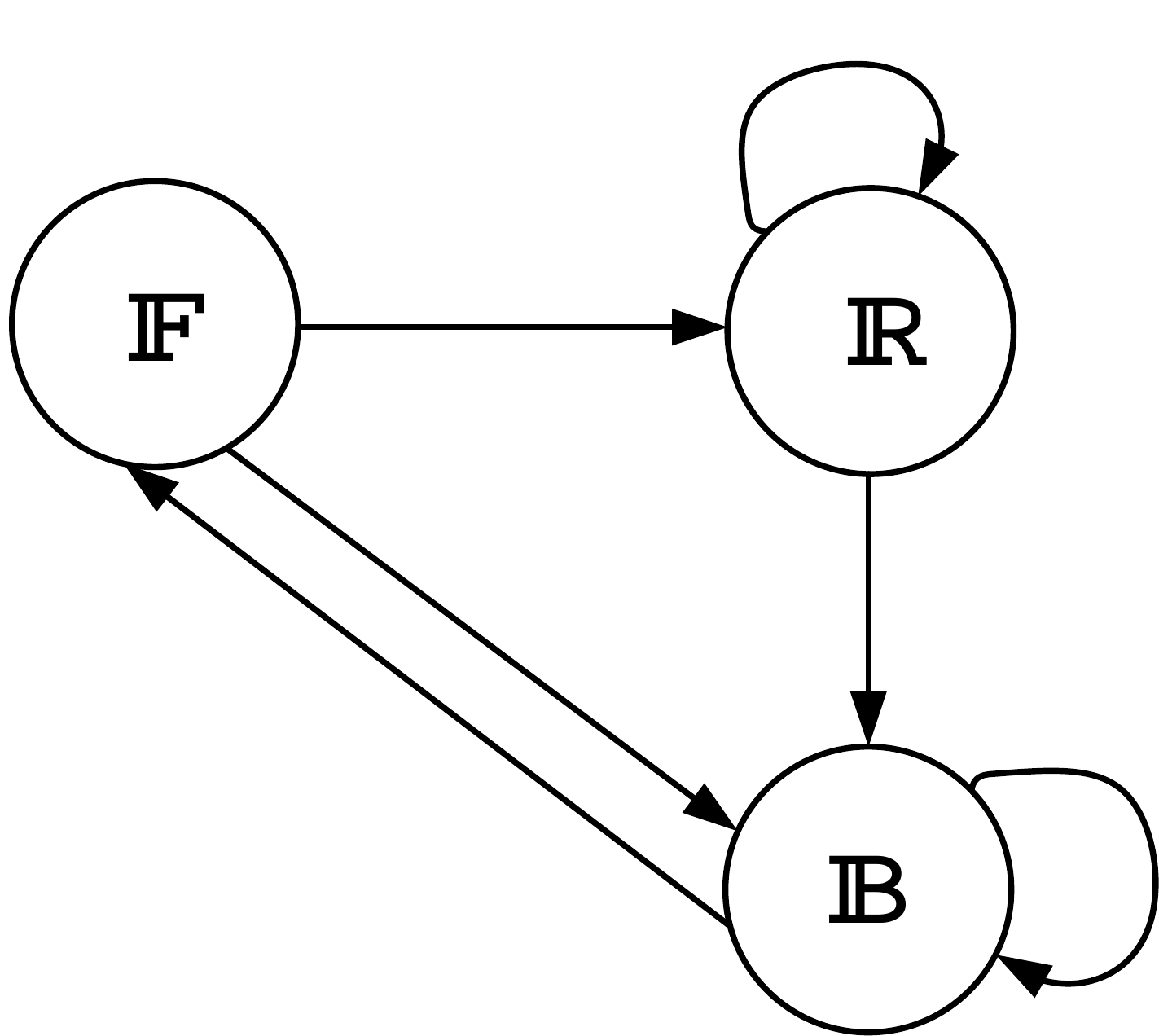}\\
  \caption{State transition diagram of Markov chain for two-hop wireless link.  Each set of states, $\mathbb{F}$, $\mathbb{R}$, and $\mathbb{B}$ contains $N$ states.}\label{fig:mvl}
\end{figure}

Suppose the current state of the system is $\mathbb F^{i}$.  It implies that the destination node $D$  has successfully received  a packet from the relay $R$ in the previous slot  and has broadcast an ACK to the sensor node $S$ and the relay node $R$. Therefore, in this current time slot, the sensor  (having received the ACK) will transmit a new packet. Suppose the position of the relay $R$ is $j$ in the next time slot. Then the system will enter the state $\mathbb R^{\,j}$  if the relay failed to decode the new packet transmitted by the sensor (hence, only  the sensor  will retransmit the packet again). On the other hand, if the relay can decode the packet successfully, then the system will enter into the state $\mathbb B^{\,j}$ (where both the relay and the sensor will transmit in the next time slot).
Similarly, if the system is in  state $\mathbb R^{i}$ currently, then its next state can only be $\mathbb R^{j}$ or $\mathbb B^{j}$. For the destination $D$ to successfully receive the packet from R, the system must be in state $\mathbb B^{i}$. When $D$ successfully receives the packet sent by the relay,   the the system will enter the state $\mathbb F^{j}$ in the next time slot. On the other hand, if the destination node cannot decode, then the system will enter the state $\mathbb B^{j}$ instead. Figure \ref{fig:mvl}  illustrates the relations between these three types of states.

\subsection{Non-threshold Communications Model}
\label{sec:nonthresh}

To compute the transition probabilities in the Markov chain, we need to know the transmission packet error rate (PER) of  the two hops   $S  \rightarrow  R$ and $R  \rightarrow  D$.
%
%This paper used a Markov chain to model the mobility of nodes and the communications process. 
Most existing work~\cite{Urgaonkar:2011:NCR:2042997.2043013}\cite{Singh2010Delay}\cite{Zhang2006Routing}  on delay tolerant networks assume a ``threshold or on/off communications model'' where data can be sent from one node to another if the two nodes are within a transmission range. Clearly, the transmission range depends on various factors including transmission power, coding and modulation scheme, channel gains, noise power and many others. Hence, each transmission scheme (and thus also the associated transmission range) governs how states in the Markov chain transit from one to another. In other words, the connection among the states are ``hardwired'' according to the chosen transmission scheme. 

Such an approach is usually simpler ``topologically'' in the underlying state transition graph. However, it is also less flexible and is very difficult (if not infeasible) to evaluate how each parameter in a transmission scheme affects the overall system performance (e.g., in terms of throughput and energy consumption).  
Our paper however assumes a ``soft communications'' model such that the choice of transmission scheme will affect the transition probabilities between states. While the state transition matrix becomes more complicated, it now becomes feasible to optimize the right parameter in a transmission scheme to optimize the system performance.

%In our numerical case study (Section \ref{sec:EvsPi}) we will adopt a specific model.  
For now, we keep a general form where the average packet error rate ($\avePER$) is a function of the hop transmission energy, $\Etl$, for transmitter $\ell=S,R$, the modulation and coding scheme (MCS), the packet length $L$,  the hop distance, $\dl$, and the one-sided additive white Gaussian noise spectral density,  $N_0$,  at the hop receiver.

\begin{equation}
\avePER_\ell= f\hspace*{-1mm} \left(\Etl,\MCS_\ell, L, \dl, N_{0} \right)
\label{eq:avePER}
\end{equation}

%Our approach has two advantages. Under the threshold communications model, the communication scheme is directly encoded into the topology of the Markov model. This makes it hard to generate the state diagram. In our model, the state diagram has more transition and thus is more connected. However, the state diagram is also more natural and symmetric. The transmission power, the modulation scheme, will only directly affect the structure of the state diagram. Instead, it only affects the transition probability of the diagram. The second advantage of this approach is that this makes it easier to see the relation or effect on the transmission power, energy throughput, or .. whatever modulation coding scheme on the system performance. This also makes optimization of the right scheme (or right power) more easily. 

%While our approach is general enough to capture a wide range of mobility or communications models, for illustration purposes, we will however assume that all the three nodes ($S$, $R$, and $D$) are collinear (i.e.\ lying on  a single straight line). The geometry of this three-node system is shown in Figure \ref{fig:systemodel}. Let $d$ be the distance between $S$ and $R$. We will  assume that at any time instance, the relay must be at a distance of $id/(N+1)$ unit from the sensor for an integer $i$ from 1 to $N$.  We will simply say that  the relay is in  position $i$.
%
%\Lcomment{END Needs to rewrite}

Let $\mathbf{M}$ be the transition matrix of our proposed Markov chain. For any states  $\mathbb C, \mathbb D \in \Omega$, we define  $\mathbf{M}(\mathbb C \rightarrow \mathbb D )$  as the transition  probability that the system is in state $\mathbb D$ in the next time slot, given that its current state is
$\mathbb C$.
%For example, $\mathbf{M}(\mathbb R^{\,i} \rightarrow \mathbb B^{\,j})$
Similarly, for any subsets $\delta, \beta \subseteq \Omega$,  $\mathbf{M(\delta\rightarrow\beta)}$ is the set of all transition probabilities $\mathbf{M(\mathbb C \rightarrow \mathbb D)}$ for $\mathbb C \in \delta$ and $\mathbb D \in \beta$. In many cases, it is instrumentally convenient to regard $\mathbf{M(\delta\rightarrow\beta)}$ as a submatrix of $M$  such that its entries in the $\mathbb C^{th}$ row and the $\mathbb D^{th}$ column is $\mathbf{M(\mathbb C \rightarrow \mathbb D)}$.

Let $\mathbb F^{1:N} $ be the set of states $\{\mathbb F^{i}, i=1, \ldots, N  \}$, and similarly we define  $\mathbb R^{1:N} $ and $\mathbb B^{1:N} $. Using our notation convention,  the transition matrix $\mathbf{M}$ can be written as in \eqref{eq:M}.
%\begin{equation}
% \textbf{M} =
% \begin{pmatrix}
%   \mathbf{0} & \mathbf{M}({\mathbb F^{1:N}\rightarrow\mathbb R^{1:N}}) & \mathbf{M}({\mathbb F^{1:N}\rightarrow\mathbb B^{1:N}})\\
%  \mathbf{0} & \mathbf{M}({\mathbb R^{1:N}\rightarrow\mathbb R^{1:N}}) & \mathbf{M}({\mathbb R^{1:N}\rightarrow\mathbb B^{1:N}}) \\
% \mathbf{M}({\mathbb B^{1:N}\rightarrow\mathbb F^{1:N}}) & \mathbf{0} & \mathbf{M}({\mathbb B^{1:N}\rightarrow\mathbb B^{1:N}}) \\
% \end{pmatrix}
%\end{equation}

%\newcounter{MYtempeqncnt}
\begin{figure*}[!t]
% ensure that we have normalsize text
\normalsize
% Store the current equation number.
%\setcounter{MYtempeqncnt}{\value{equation}}
% Set the equation number to one less than the one
% desired for the first equation here.
% The value here will have to changed if equations
% are added or removed prior to the place these
% equations are referenced in the main text.
\begin{equation}
 \textbf{M} =
 \begin{pmatrix}
   \mathbf{0} & \mathbf{M}({\mathbb F^{1:N}\rightarrow\mathbb R^{1:N}}) & \mathbf{M}({\mathbb F^{1:N}\rightarrow\mathbb B^{1:N}})\\
  \mathbf{0} & \mathbf{M}({\mathbb R^{1:N}\rightarrow\mathbb R^{1:N}}) & \mathbf{M}({\mathbb R^{1:N}\rightarrow\mathbb B^{1:N}}) \\
 \mathbf{M}({\mathbb B^{1:N}\rightarrow\mathbb F^{1:N}}) & \mathbf{0} & \mathbf{M}({\mathbb B^{1:N}\rightarrow\mathbb B^{1:N}}) \\
 \end{pmatrix}\label{eq:M}
\end{equation}
% Restore the current equation number.
%\setcounter{equation}{\value{MYtempeqncnt}}
% IEEE uses as a separator
\hrulefill
% The spacer can be tweaked to stop underfull vboxes.
\vspace*{4pt}
\end{figure*}

%Note that some of the entries in $ \textbf{M} $ is zero. For instance,
%$\mathbf{M(\mathbb F^{i} \rightarrow \mathbb F^{j})} =0$ .

We would also like to point out the following assumption in our model. The probability that the relay $R$ can decode a packet sent by the sensor $S$ depends only on the distance between $S$ and $R$ (or equivalently, the current position $i$ of $R$). This probability is independent of the position of the relay ($j$) in the next time slot.  Likewise, the probability that $D$ can decode a packet sent by $R$ depends only on the distance between $R$ and $D$.

The set of probabilities needed for the transition matrix is:
\begin{eqnarray}
&& \hspace{-0.7cm} \text{M}\hspace*{-1mm}\left(\mathbb{F}^i \rightarrow \mathbb{R}^j \right) \nonumber \\ 
&& \hspace{-0.2cm}  = \f{\EtS,\MCS_S,L,\dS, N_{0}} P^{s}[i,j] \label{eq:5}\\ 
&& \hspace{-0.7cm} \text{M}\hspace*{-1mm}\left(\mathbb{F}^i \rightarrow \mathbb{B}^j \right) \nonumber \\ 
&& \hspace{-0.2cm}  =  \left(1-\f{\EtS,\MCS_S,L,\dS, N_{0}}\right)   P^{s}[i,j] \\
&& \hspace{-0.7cm} \text{M}\hspace*{-1mm}\left(\mathbb{R}^i \rightarrow \mathbb{R}^j \right) \nonumber \\ 
&& \hspace{-0.2cm}  = \f{\EtS,\MCS_S,L,\dS, N_{0}}  P^{s}[i,j]\\
&& \hspace{-0.7cm} \text{M}\hspace*{-1mm}\left(\mathbb{R}^i \rightarrow \mathbb{B}^j \right) \nonumber \\ 
&& \hspace{-0.2cm}  =  \left(1-\f{\EtS,\MCS_S,L,\dS, N_{0}}\right)  P^{s}[i,j]\\
&& \hspace{-0.7cm}\text{M}\hspace*{-1mm}\left(\mathbb{B}^i \rightarrow \mathbb{B}^j \right) \nonumber \\ 
&& \hspace{-0.2cm}  =  \f{\EtR,\MCS_R,L,\dR, N_{0}} P^{s}[i,j]\\
&& \hspace{-0.7cm} \text{M}\hspace*{-1mm}\left(\mathbb{B}^i \rightarrow \mathbb{F}^j \right) \nonumber \\ 
&& \hspace{-0.2cm}  =   \left(1-\f{\EtR,\MCS_R,L,\dR, N_{0}}\right)  P^{s}[i,j]. \label{eq:10}
\end{eqnarray}

%% REMOVED THIS ... USED INDEX k ELSEWHERE!
%Based on this assumption, we can show that
%\begin{equation}
%\frac{\mathbf{M}({\mathbb F^{i}\rightarrow\mathbb R^{j}})}{\mathbf{M}({\mathbb F^{i}\rightarrow\mathbb R^{k}})}
%=
%\frac{\Pr(i \to j)}{\Pr( i \to k)}
%\end{equation}
%where
%$\Pr({i}\rightarrow {j})$ (or $\Pr({i}\rightarrow {k})$) is a the probability that the relay will be in the position $j$
%(or $k$) in the next time slot, given its current position is $i$, as defined in equation (\ref{eq:pij}).

Having modelled the communications process by a Markov chain, we can
calculate the limiting probability of each state by solving a system of linear equations~\cite{ross2007introduction}.
Specifically, let $\pi = (\pi_{\mathbb A} , \mathbb A \in \Omega)$ be the vector of limiting probabilities. Then
\begin{equation}
\label{eq:steadystateeqn}
    \mathbf{\pi}  \, \mathbf{M}= \mathbf{\pi}
\end{equation}
and $\sum_{\mathbb A \in \Omega} \pi_{\mathbb A}  = 1$. After computing these limiting probabilities\footnote{
The complexity to determine the limiting probabilities is of order ${\mathcal O}(M^{3})$, where $M$ is  the size of the Markov chain state-space. If one would plot the performance curve with $K$ grid points, then the overall complexity will be of order  ${\mathcal O}(M^{3}K)$.
}, we can then evaluate the system throughput and the average packet transmission energy.
In particular, the throughput (denoted by $\pi_{0}$) is given as follows:
\begin{equation}
\pi_{0}=\sum^{N}_{i=1} \pi_{\mathbb F^{i}}.
\end{equation}
Roughly speaking, if the system has been operated for $T$ time slots (where $T$ is a very large number), then by Little's law \cite{queueing1971}, the expected number of packets that have been transmitted successfully is given by $T \pi_{0}$. Equivalently, the expected packet transmission delay  $\tau$ is $ {1}/{\pi_0}$ slots.
The expected packet transmission energy consumed by the sensor source (denoted by $\Etot$) is thus given by
\begin{equation}\label{eq:Emodel}
   \Etot=   \frac{\EtS \, L}{\pi_0}.
\end{equation}

Before we end this section, we would also like to highlight that this paper's objective is not to derive the most detailed and accurate model for wireless sensor networks. Instead, we aim to derive insights and system design criteria or principals from our analysis, realising that insights obtained from our simpler model will still be valid in more sophisticated models. As an example, one idea obtained and verified from our model is that time-sharing can improve system performance (see Section \ref{sec:MCTS}).

\subsection{Extensions}
\label{sec:ext}

The Markov chain formulation has been used to evaluate the energy-performance trade-off for a two-hop network. This approach really is quite general and could readily be applied to more sophisticated network models, especially where the Markov chain and calculation of its transition matrix can be automated.  
Some possibilities include larger, two-dimensional networks; nodes with finite buffers; multiple relays; multiple antennas for multiplexing or beamforming; networks with interference; adaptive modulation and coding schemes; and, variable packet length protocols. Through modification of the definition of state, the Markov chain can be used to model the behavior of the network in terms of a variety of parameters. In addition to transmission energy and expected throughput, the state could be defined so as to investigate latency, buffer overflow, battery depletion and so on. The Markov chain models similarly could be used to simulate time evolving performance of such networks.

For example, in our proposed model, we assume no interaction between sources and relays. Therefore, the source will continue transmitting the data, even if the relay has already received the data packet. Suppose that  there is only one relay and that the relay can send an ACK back to the source node. Then the transmission scheme will be more energy efficient if the source can stop its transmission whenever an ACK has been received from the relay. Our Markov model can easily be adapted to take into account the changes. Another extension is by relaxing the constraint that the sink node can send an instantaneous ACK to  all sensor and relay nodes. We can instead requiring the ACK be sent from the sink node to the source node via the two-hop link, or with some tangible delay.

The exposition for the energy-throughput trade-off presented in this paper focuses on a total energy cost only for the sensor source transmission, $\Etot$.  Another extension of our model is to change this total cost function.   So long as the total cost can be calculated in terms of the expected amount of time the finite state machine spends in any given state, a similar approach to analysis can be taken.
 
%It is worth mentioning however that when there are multiple relays, it may be more efficient if the source can continue transmission ensuring more relays received the data packets and can thus forward it to the sink node. 

%\subsubsection{Two-hop routing}

While this paper considered only the single relay scenario, our proposed approach can also be naturally extended to cases where there are multiple relays.  In such a scenario, the infostation can receive data via one of the relays. As before,  we can obtain the Markov state transition diagram in Figure \ref{fig:mvl} first and evaluate the transition probabilities according to the chosen PHY and MAC model. Then, the trade-off between delays and throughput can be evaluated by finding the stationary distribution of the associated Markov chain. 
This approach however has a downside that the complexity to find the stationary probability of the Markov chain scales with the number of relays. In the context of two-hop routing, some tricks can be employed to simplify the analysis.  

Under two-hop routing, data are sent from the source sensor $S$ to the sink $D$ via multiple relays. As packets are not sent among relays, the underlying network is essentially a set of parallel two-hop links connecting the source to the sink. Furthermore, the ``transmission process'' of these two-hop links are essentially independent\footnote{
The independence assumption holds in the scenario being considered when there is only a single source and a single destination, or  when the relays have sufficiently large buffers. However, it does not hold in the general sense. When there are multiple sources and the relays have limited buffers, one will need to extend the Markov chain model to capture the interactions among nodes. 
}. 

Let $\eta$ denote the random time required to send a data packet successfully from the source $S$ to the sink $D$ via a single two-hop link. Suppose there are $m$ relays (or equivalently two-hop links). Let 
$
\eta_{k}
$
be the time required to send a data packet from the source to the sink relay $k$, for $k=1, \ldots, m$. Then
$\{\eta_{1}, \ldots, \eta_{m}\}$ is a collection of i.i.d. random variables, each of which has the same distribution as the expected delay, $\tau$, for the single relay case. 
From the sink node's perspective, a packet has been received at time $t$ if and only if 
\begin{equation}
t \ge \min(\eta_{1}, \ldots, \eta_{m}).
\end{equation}
By analyzing the distribution of $\eta$, we can also characterise the time required to transmit a packet via $m$ relays. For example, let $p(t)$ be the probability that $\Pr(\eta \le t)$ and $m$ (i.e., the number of relays) be a poisson random variable with mean $\lambda$. Then 
\begin{equation}
\Pr( t \ge \min(\eta_{1}, \ldots, \eta_{m})  ) = 1 - e^{-\lambda p(t)}.
\end{equation}

%Previously we presented our model for simplicity based on $S$, $R$ and $D$ being collinear  (i.e.~lying on a straight line).
%
%For the collinear case, we let there be~$N$ possible relay positions between the source $S$ at location $i=0$ and the sink $D$ at $i=(N+1)$.  We let the distance between~$S$ and~$D$ be~$d$, and denoted the distance $S$ to $R$  by $d_S$, and the distance $R$ to $D$ by $d_R$.  
%  For the relay in position~$i$:
%\begin{eqnarray}
%d_S&=& id/(N+1) \nonumber \\
%       &=& i \Delta\\
%d_R&=& d - d_S \nonumber\\
%       &=& (1 - i/(N+1))d \nonumber\\
%       &=& (N+1 -i)\Delta
%\end{eqnarray}
%where we have introduced the distance between successive positions $\Delta = d / (N+1)$.
%
%\begin{figure}[ht]
%\centering
%  % Requires \usepackage{graphicx}
%  \includegraphics[scale=1.4]{systemodel.pdf}\\
%  \caption{1D link geometry for $S$ = sensor, $R$ = relay, and $D$ = destination nodes.  There are $N$ possible positions for the relay, current position, $i$.}\label{fig:systemodel}
%\end{figure}
%
%
%%\begin{figure}[ht]
%%\centering
%%  % Requires \usepackage{graphicx}
%%  \includegraphics[scale=1.40]{brownianmotion.pdf}\\
%%  \caption{Bounded Brownian motion of the relay node R with $s$ random walk steps taken between clock ticks $t$ and $t+1$.}\label{fig:brownian}
%%\end{figure}
%
%Equation (1), page 7 of paper describes the symmetric random walk.

In addition to extension to two-hop routing networks, we can also extend by incorporating a more sophisticated mobility model. For simplicity, our previous model assumed a ``collinear mobility model'' in which the relay $R$ moves along the straight line between the source $S$ and the destination $D$. It is worth noticing that the geometry between $S$, $R$ and $D$ is not important in our model. Instead, only the relative distances among the nodes matter. Therefore, it  is in fact quite  straightforward to extend our approach using a  two-dimensional mobility model. In the following, we will sketch the ideas how it can be done.

\begin{figure}[ht]
\centering
  \includegraphics[width=1\columnwidth]{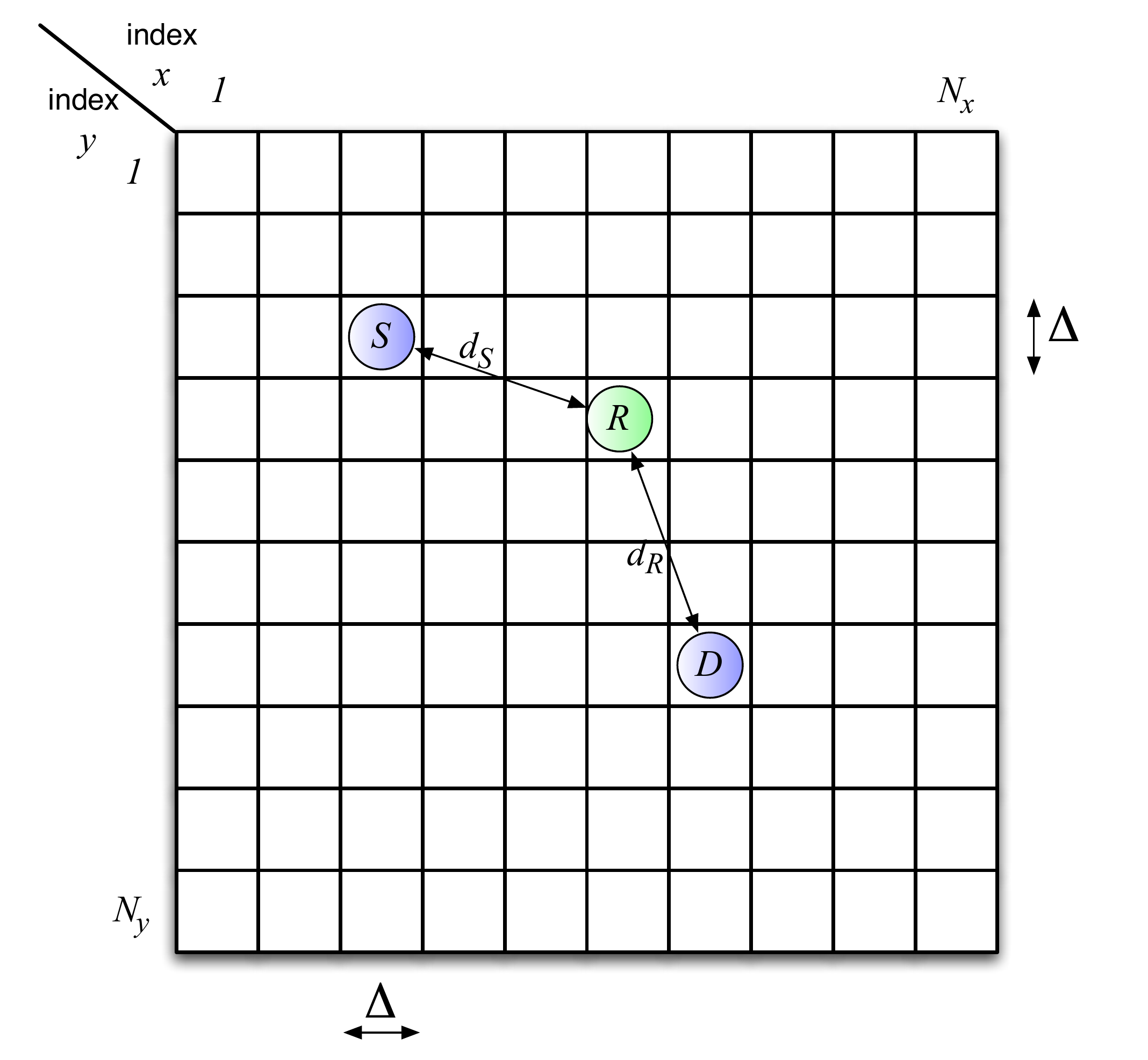}\\
  \caption{2D link geometry for $S$ = sensor, $R$ = relay, and $D$ = destination nodes.  The relay can be in one of the $N_x N_y$ possible positions indexed  by $i$ or equivalently $(x_{i},y_{i})$.}\label{fig:2Dfield}
\end{figure}

 We assume a $N_x \times N_y$ grid, with scale $\Delta$ as shown in Figure~\ref{fig:2Dfield}. There are $N_x$ columns and $N_y$ rows.  The position of the sensor $S$ and the destination $D$ are arbitrary, but fixed.  The relay $R$ can move about the 2D field.  For position $i = 1, \ldots, N_x N_y$, we have
\begin{eqnarray}
x_i &=& \left(\left(i-1\right) \mod N_x\right) + 1\\
y_i &=& \lceil i/N_x \rceil
\end{eqnarray}

\begin{figure}[ht]
\centering
  \includegraphics[width=0.3\columnwidth]{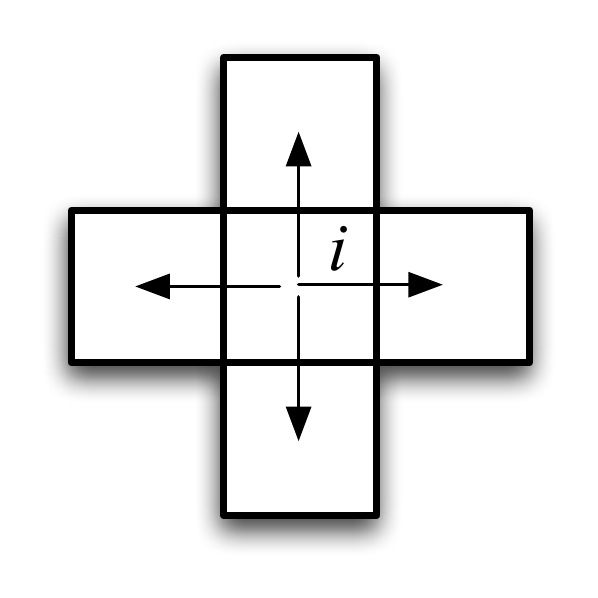}\\
  \includegraphics[width=1\columnwidth]{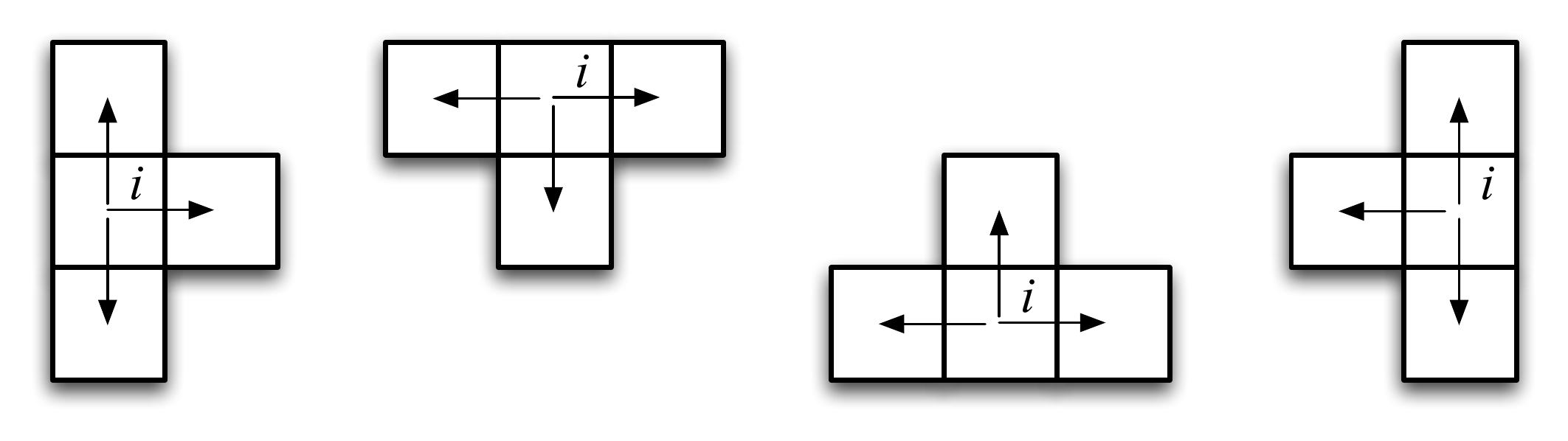}
  \includegraphics[width=1\columnwidth]{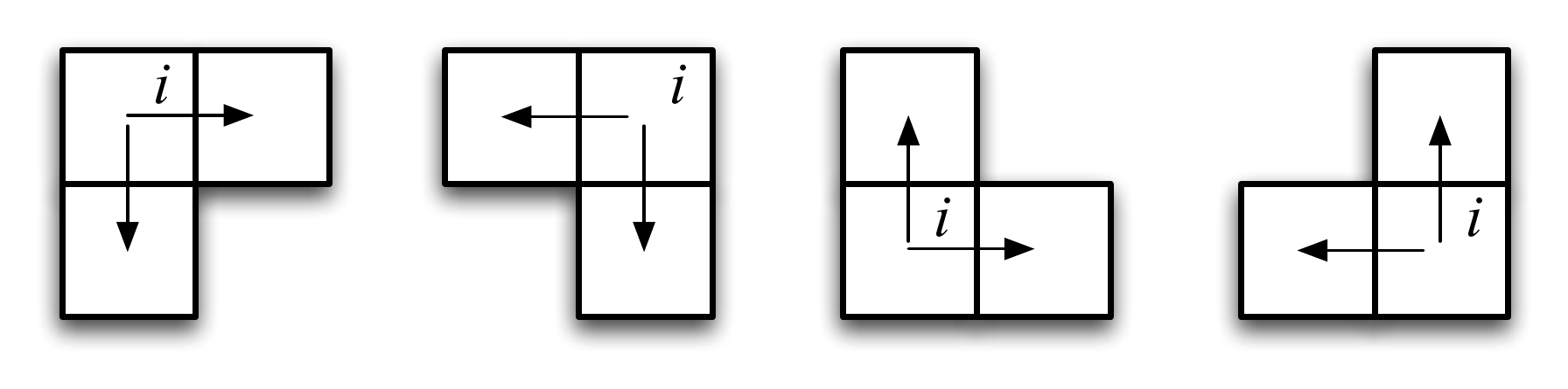}
  \caption{2D random walk cases for the relay in position $i$.}
  \label{fig:2Dcases}
\end{figure}

For the relay in position $i$, the relevant distances are:
\begin{eqnarray}
d_S &=& \Delta\sqrt{(x_i - x_S)^2 + (y_i - y_S)^2}\\\
d_R &=&  \Delta\sqrt{(x_i - x_D)^2 + (y_i - y_D)^2}
\end{eqnarray}
In this 2D grid, a symmetric random walk can be established.  One possibility is to replace \eqref{eq:1} with:
\begin{equation}
P[i,j] = \left\{ 
\begin{array}{ll}
1 - p_{\text{move}}			& x_i = x_j \text{ and } y_i = y_j\\
p_{\text{move}}  / 4			& |x_i - x_j| = 1 \text{ xor } |y_i - y_j| = 1\\
                                   			& 1 < x_i < N_x, 1< y_i < N_y\\
p_{\text{move}}  / 3			& |x_i - x_j| = 1 \text{ xor } |y_i - y_j| = 1\\
						&  x_i \in \{1, N_x\} \text{ xor }  y_i \in\{1, N_y\} \\
p_{\text{move}}  / 2			&  |x_i - x_j| = 1 \text{ xor } |y_i - y_j| = 1\\
						&  x_i \in \{1, N_x\} \text{ and }  y_i \in\{1, N_y\} \\
 0                                  			& \text{otherwise.}
\end{array}
\right.
\end{equation}
These equations are based on consideration of Figure~\ref{fig:2Dcases}.
In this case, the state space is now dimensioned by $N = N_x N_y$ and \eqref{eq:5} -- \eqref{eq:10} apply as before.

%%%%%%%%%%%%%%%%%%%%%%%%%%%%%%%%%%%%%%%%%%%%%%
\section{Energy-Throughput Trade-off}
\label{sec:char}
\subsection{Trade-off Optimization}
\label{sec:opt}

Using the Markov chain formulation described in Section \ref{sec:system}, we can now characterize the trade-off between the packet transmission energy (consumed by the sensor node) and the transmission throughput in our mobile relay model for a numerical case study.

In \cite{zhu2011trade}, a similar trade-off between packet transmission energy  and throughput was investigated.  The model proposed here has  two important differences. First,  the relay $R$ in  \cite{zhu2011trade} was \emph{stationary} and did not move during the whole communication process.  Second, in that earlier work, the transmission energy consumption included  the energies consumed by both the sensor node $S$ and the relay node $R$ for the case where $\EtS = \EtR = \Et$.  It was shown in that paper that the packet transmission energy and the  network throughput depend highly on both the transmission symbol energy level $\Et$  and the position of the stationary relay.
In this paper, we focus on the trade-off between the packet transmission energy consumed by the sensor node $S$ and the throughput when the relay is mobile.

In the following, we will use our proposed Markov chain model to analyze an example to
 better understand this throughput-energy trade-off.
In our example, the distance between $S$ and $D$ is $d=300$m, which is the same as that used in our earlier setup  in \cite{zhu2011trade}. 
For the average packet rate ($\overline{\text{PER}}$)  of equation (\ref{eq:avePER}), we adopt the  model proposed in  \cite{zhou2008energy} (based on \cite{liu2004cross})  for uncoded packet over a Rayleigh fading channel:
\begin{align}
\overline{\text{PER}}  & =\frac{a_{n}r^\alpha N_{0} }{ r^\alpha N_{0}+g_{n} \Et G}e^{-\gamma_{pn}(g_{n}  +\frac{N_{0}r^\alpha}{ \Et G})} \nonumber \\
& \qquad \qquad +(1-e^{\frac{-\gamma_{pn} N_{0} r^\alpha}{ \Et G}})
\label{eq:avePERRayleigh}
\end{align}
where we have dropped the implied subscript $\ell$ designating the transmitter $S$ or $R$.  The form of this equation comes from the averaging over flat-fading Rayleigh channels, and an approximate exponential expression for the PER for an additive white Gaussian noise channel at a given SNR.   Details are in \cite{zhou2008energy}.  We use parameters from the model in \cite{liu2004cross} with $b=1$ for BPSK, antenna gain $G=1$, and with parameters $a_{n}=67.7328$, $g_n=0.9819$, $r_{pn}=6.3281$ dB, that depend on the MCS and the packet length, chosen to be $L=1080$.   We assume  that $N=10$  (i.e.\ the relay can be in any one of ten possible positions).
The level of symbol energy $\Etl$ (for $\ell = S, R$) can take any value from $ 1\times 10^{-7}$ J to $2\times 10^{-4}$ J. In addition, $p_{\text{move}}$ is chosen to be $0.98$.

  \begin{figure*}[ht]
\centering
  \includegraphics[scale=0.7]{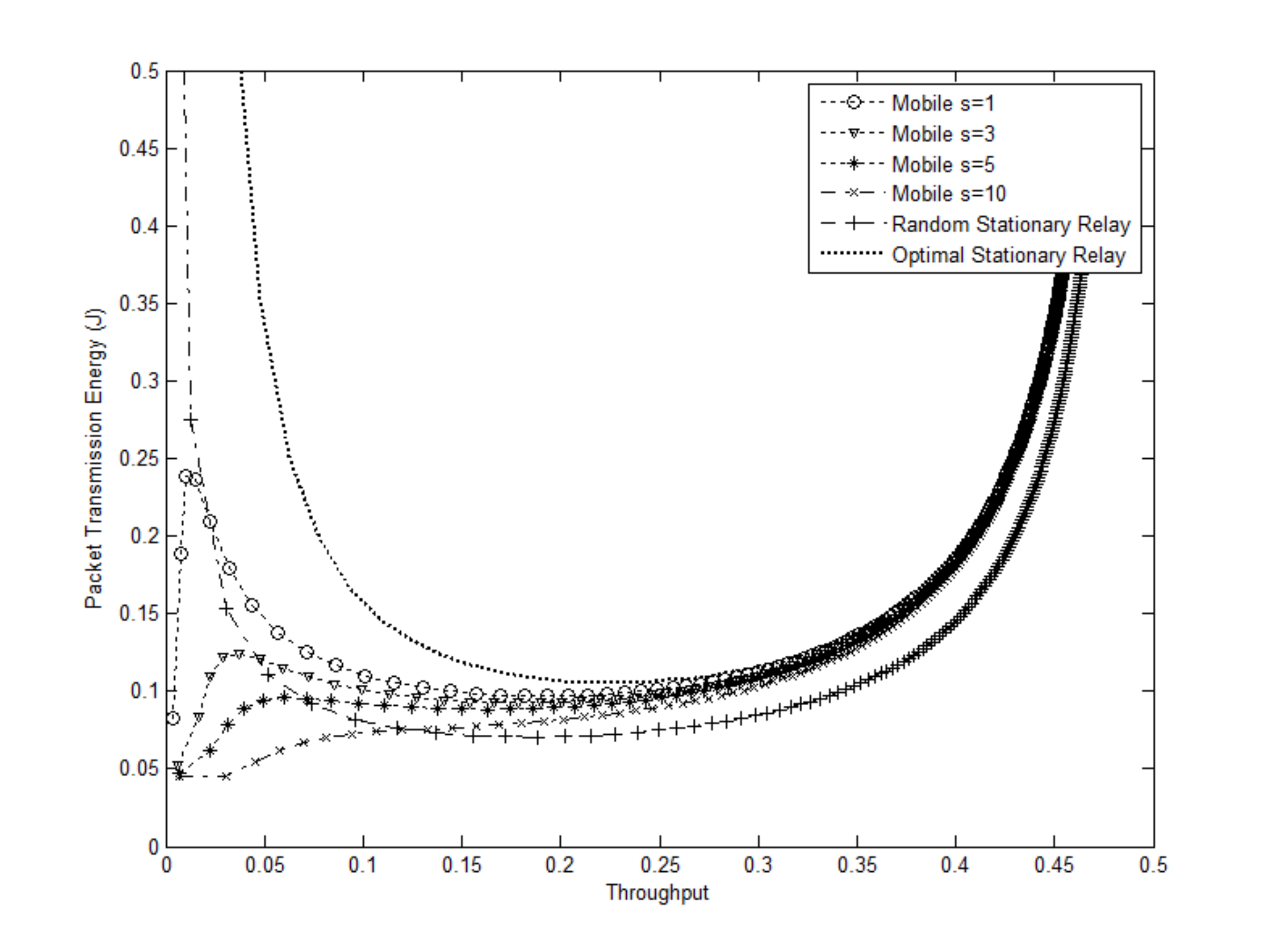}\\
  \caption{Trade-off in total energy and throughput for mobile relay link.} \label{fig:mov&jump}
\end{figure*}

Figure \ref{fig:mov&jump} depicts the throughput-energy trade-off for various mobility speeds  $s$ (i.e.\ the number of steps that a relay can move in any one time slot). For each mobility speed $s$ and a chosen symbol energy level $\EtS = \EtR = \Et$, we can evaluate a  throughput-energy
tuple $(\pi_{0}, \Etot)$ by computing the limiting probabilities of the states in the Markov chain. By varying $\Et$, we thus obtain a throughput-energy curve for a given mobility speed $s$.
It is not difficult to see that if $\Et$ increases, then $\pi_{0}$ also increases. However, increasing $\Et$ does not directly imply that the packet transmission energy $\Etot$  also increases.
We see in Figure \ref{fig:mov&jump}, especially for $s=1$, that $\Etot$ may decrease even if $\Et$ increases. Such a reduction in $\Etot$ is not hard to explain.  While  the increase of  $\Et$ requires an increase in the amount of energy needed to transmit a packet in one time slot, it also increases the opportunities for a packet to be received successfully by the relay or by the destination node. Therefore, the average number of retransmissions of a packet   can also be reduced. Our simulation shows that this reduction in the number of retransmissions can result in a substantial reduction in the  total energy needed to transmit a packet.

 % comparison to stationary
One important conclusion we can draw from the mobility results in Figure~\ref{fig:mov&jump}  is that relay mobility can help reducing packet
transmission energy. To justify our claim, we will compare our obtained throughput-energy curves for various mobility speeds with the ones for when the relay is stationary. In particular, we consider two stationary relays scenarios. In the first scenario (called the \emph{random stationary relay scenario}), the position of the relay is randomly chosen before the communications process. Once the position is selected, the relay will stay in the same position during the whole process.  For fair comparison, the probability that the stationary relay will be at the position $i$ will be the same as the probability for that the mobile relay will be at the same position.

 \begin{figure*}[ht]
\centering
  \includegraphics[scale=1]{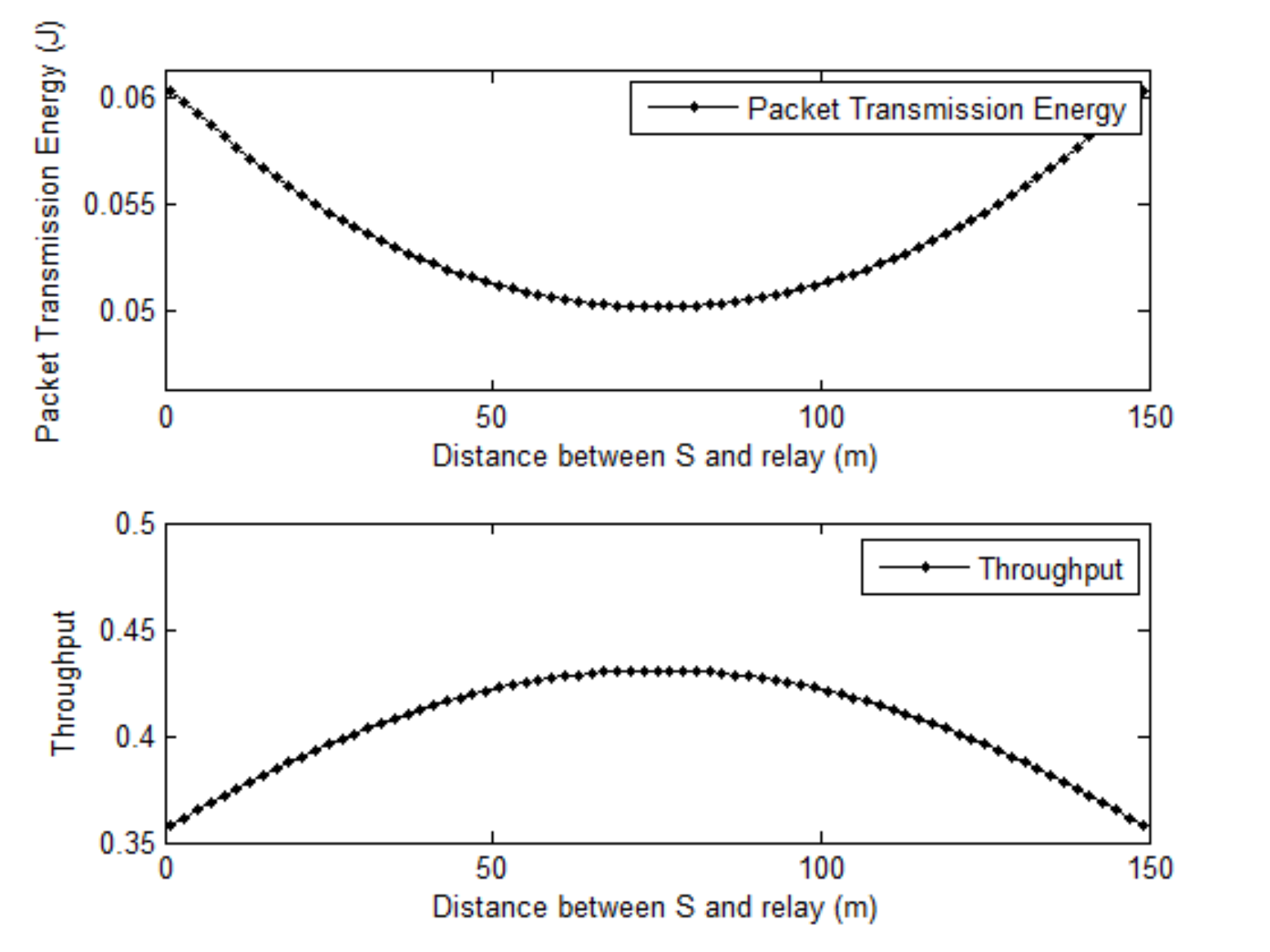}\\
  \caption{Energy transmission of $S$ and $R$ fixed at $\EtS=\EtR-\Et$.   Source energy consumption $\Etot$ as a function of relay position in meters.} \label{fig:fixedEt_adjustposition}
\end{figure*}

\begin{figure*}[ht]
\centering
  \includegraphics[scale=1]{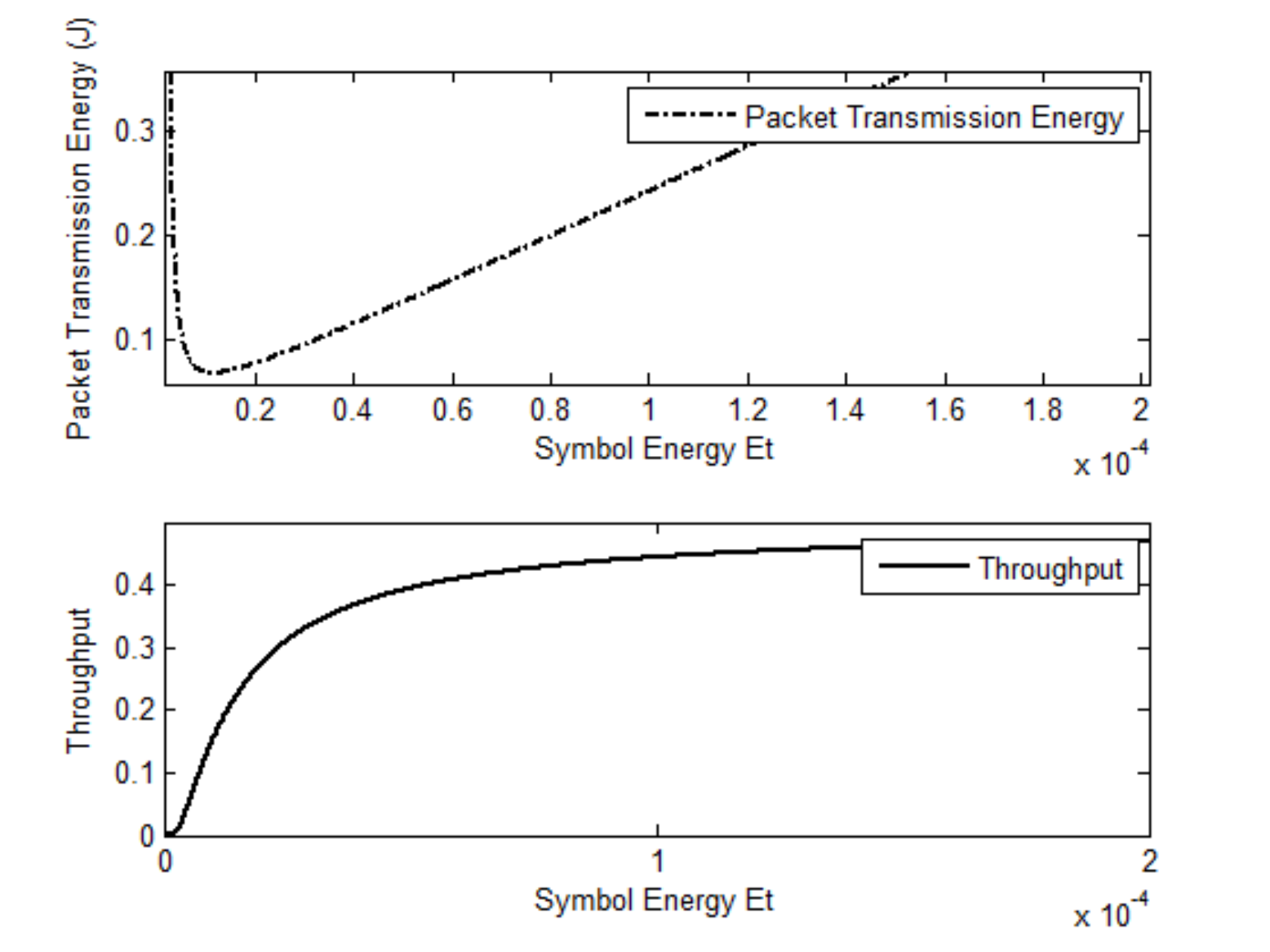}\\
  \caption{Relay node fixed at midpoint between $S$ and $D$.  Energy consumption $\EtS$ as a function of transmission energy $\Et$.} \label{fig:fixedcentre_adjustEt}
\end{figure*}

In the second scenario (called the \emph{optimal stationary relay scenario}), the position of the relay will be selected optimally. Figure \ref{fig:fixedEt_adjustposition} shows the packet transmission energy $\Etot$ consumed by the sensor source node $S$ and the throughput of the network, $\pi_0$ as a function of the relay position for a fixed level of $\Et$. We can see that  packet transmission energy and throughput are optimized when  the relay is at the midpoint between $S$ and D.
This occurs because of our assumption that both $S$ and $R$  use the same transmission energy level. Therefore, in this second scenario, we will always assume that the relay $R$ is at the midpoint between $S$ and $D$. Under this assumption, we can show the throughput-energy curve by varying the symbol energy level $\Et$. See Figure \ref{fig:fixedcentre_adjustEt}.

If we compare the throughput-energy curves with a mobile relay to the one with a random stationary relay,  we can see from Figure \ref{fig:mov&jump} that relay mobility does reduce energy consumption for the whole range of throughputs.
 In fact, the larger  the mobility speed $s$, the lower the packet transmission energy $\Etot$. We also observe that the reduction in transmission energy from mobility decreases as the throughput increases. This phenomenon can be explained as follows.  When the system is operated in the high throughput regime, packets must reach the destination in a fairly short period of time. Clearly, within such a short time, the location of the relay will not change significantly. Thus, the relay looks as if it is stationary. Therefore, we do not expect that mobility can greatly reduce transmission energy in the high throughput regime.

However, when comparing with the throughput-energy curve in the optimal stationary relay scenario, we notice seemingly contradictory outcomes that for a wide range of throughput levels, a stationary relay performs better than a mobile relay. The reason is largely due to that the position of the stationary relay is optimally chosen, while the position of mobile relay changes continuously.  Therefore, in the high throughput regime (where mobility does not help in general), an optimally positioned relay performs better than a mobile relay in the sense of reducing packet transmission energy.  However, if we focus only on the low throughput regime, relay mobility can still reduce transmission energy, as illustrated from Figure  \ref{fig:mov&jump}.

\subsection{Enhancement by Time-Sharing}
\label{sec:MCTS}

There are scenarios where energy is critically scarce (especially when it is essential to lengthen the lifespan of a battery-operated sensor network). The numerical example in the earlier section showed a very interesting property: Suppose that the relay is stationary (or is moving very slowly) and that the current level of symbol energy $\Et$ is very small. If 
$\Et$ increases slightly, then the throughput may only increase slightly while the transmission energy of each packet may increase significantly.
In fact, our example showed that it is sometimes better to operate the system at an even higher symbol energy level to reduce packet transmission energy (while at the same time also increase the network throughput).   In this section, we will propose a ``time-sharing'' approach to further reduce packet transmission energy.

The idea is very simple. Instead of using only one symbol energy level, we allow the use of two levels, $E^{\text{Tx}}_{\,\text{low}}$ and $E^{\text{Tx}}_{\,\text{high}}$. For each packet, when it is first transmitted, the sensor node will randomly select whether the packet should be sent using the
symbol energy level $E^{\text{Tx}}_{\,\text{low}}$ (with a probability $1-q$) or $E^{\text{Tx}}_{\,\text{high}}$ (with a probability $q$). Once the level is determined, the packet will be transmitted using the same symbol energy level for all subsequent retransmissions.
It turns out that this time-sharing approach can reduce packet transmission energy and also offer an opportunity to provide priority service to transmit delay-sensitive packets.

\begin{figure}[ht]
\centering
  \includegraphics[scale=0.5]{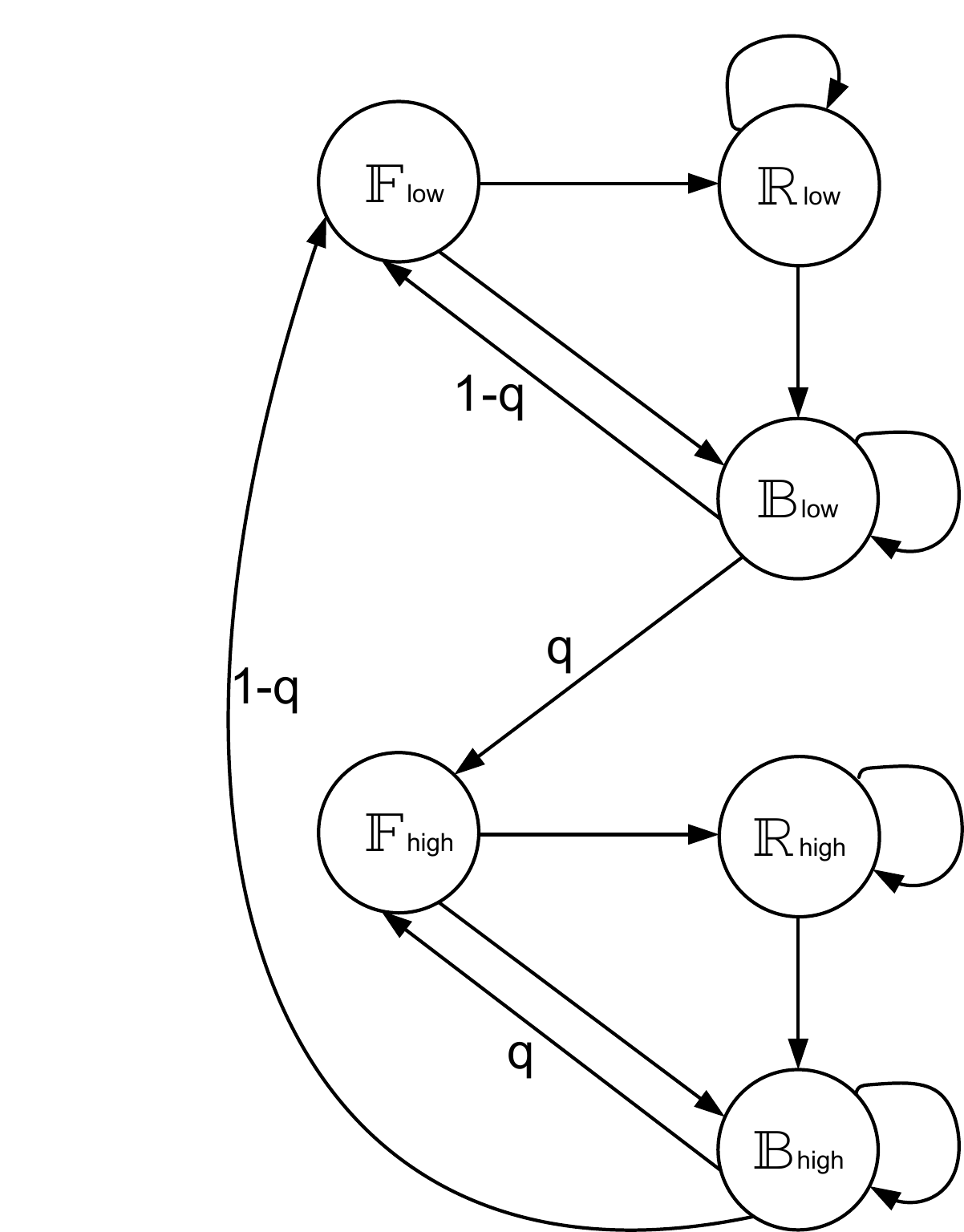}\\
  \caption{State transition diagram of time-sharing scheme.}\label{fig:mv2}
\end{figure}

As before, we will model the whole communications process with a Markov chain, which consists of  the following states
\begin{equation}
\Omega^{*} \triangleq \left\{ {\mathbb F^{i}_{\text{low}}},  {\mathbb F^{i}_{\text{high}}}, {\mathbb R^{i}_{\text{low}}},  {\mathbb R^{i}_{\text{high}}} ,  {\mathbb B^{i}_{\text{low}}} ,  {\mathbb B^{i}_{\text{high}}}, \: i=1,\ldots, N \right\}.
\end{equation}
The definition of the states are defined similarly as before.
For example, the communication system is in state ${\mathbb F^{i}_{\text{high}}}$ if 1) the sensor node is going to transmit a new packet, 2) the position of the relay is $i$, and 3) the packet will be transmitted by the sensor with a symbol energy $E_{\, \text{high}}$.
Other states can also be defined similarly as before. Figure \ref{fig:mv2} illustrates the relations between these six types of states.

Let $\pi^{*} = (\pi^{*}_{\mathbb A}, \mathbb A \in \Omega^{*})$ be the vector of limiting probabilities. Note that these limiting probabilities depend on the value of the time-sharing factor $q$. The  transmission throughput $\pi^{*}_{0}$ can be computed by
\begin{equation}
\pi^{*}_{0} = \sum _{i=1}^{N}\pi^{*}_{\mathbb F^{i}_{\text{low}}}+ \sum_{i=1}^{N}\pi^{*}_{\mathbb F^{i}_{\text{high}}}.
\end{equation}

Average packet transmission energy can also be computed similarly.
Let
\begin{align}
\pi^{*}_{\text{low},0} &= \sum_{i=1}^{N} \pi^{*}_{\mathbb F^{i}_{\text{low}}} \\
\pi^{*}_{\text{low}} & =\sum_{i=1}^{N} \pi^{*}_{\mathbb F^{i}_{\text{low}}}+ \sum_{i=1}^{N} \pi^{*}_{\mathbb R^{i}_{\text{low}}}+\sum_{i=1}^{N} \pi^{*}_{\mathbb B^{i}_{\text{low}}} \\
\pi^{*}_{\text{high},0} &={ \sum_{i=1}^{N} \pi^{*}_{\mathbb F^{i}_{\text{high}}} } \\
\pi^{*}_{\text{high}} & =\sum_{i=1}^{N} \pi^{*}_{\mathbb F^{i}_{\text{high}}}+\sum_{i=1}^{N} \pi^{*}_{\mathbb R^{i}_{\text{high}}}+\sum_{i=1}^{N} \pi^{*}_{\mathbb B^{i}_{\text{high}}}.
\end{align}

Then the average packet transmission energy is given by
\begin{eqnarray}\label{eq:totale}
 E^{*\text{total}}_{S} &=& L \left(\frac{E^{\text{Tx}}_{\text{low}}\, \pi^{*}_{\text{low}} + E^{\text{Tx}}_{\text{high}}\, \pi^{*}_{\text{high}}   }{ \pi^{*}_{0}}\right)
\end{eqnarray}

We now evaluate the performance of our time-sharing approach by considering a similar setup to Section \ref{sec:char}: Let $s=1$ and $N=10$ so that the relay can be at only one of the 10 possible positions between $S$ and $D$. We let $E^{\text{Tx}}_\text{low} = 1.0\times10^{-7}$ J and $E^{\text{Tx}}_\text{high} = 1.8\times10^{-5}$ J.
For each choice of the time-sharing parameter $q$ (from 0 to 1), we can compute the
tuple $( \pi^{*}_{0}, E^{*\text{total}}_{S}) $. By varying $q$, we obtain the throughput-energy curve formed by connecting all the tuples $( \pi^{*}_{0}, E^{*}_{\rm total}) $.
As transmission delay $\tau^*$ is equal to the reciprocal of throughput, if we are interested,  we can also connect all the tuples $( 1/\pi^{*}_{0}, E^{*\text{total}}_{S}) $ to form the delay-energy curve.
 In fact, in some cases, it will be more convenient to examine  the delay-energy curve directly.

\begin{figure*}[ht]
\centering
  \includegraphics[scale=0.7]{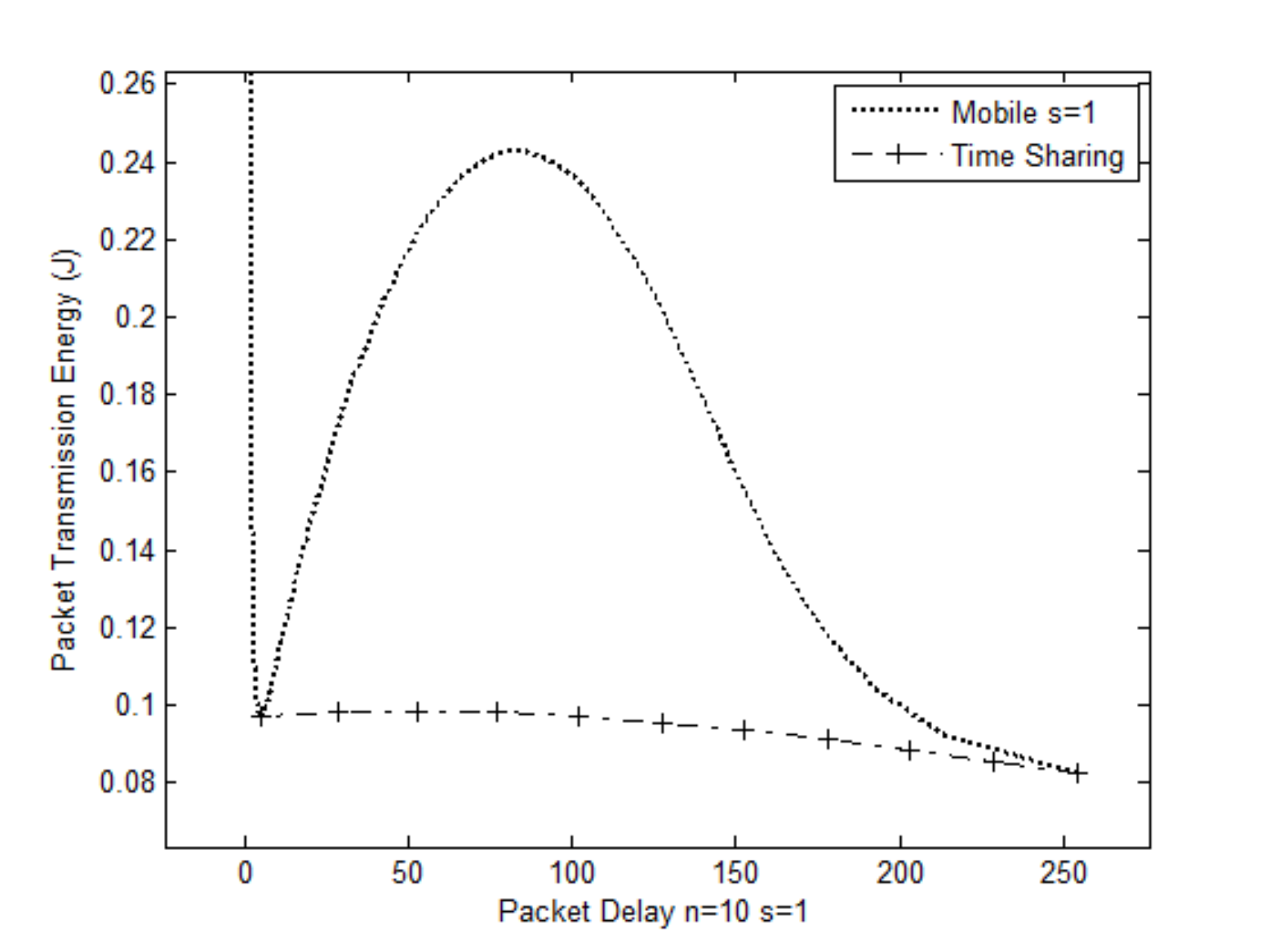}\\
  \caption{Total energy consumption for sensor node versus average packet delay (1/throughput).}\label{fig:timesharen10s1a}
\end{figure*}

In Figure \ref{fig:timesharen10s1a}, we show the delay-energy curves obtained from the time-sharing approach. And for comparison, we also plot the delay-energy curve of the system we originally proposed in Section \ref{sec:char}. If we compare the two curves, we can see that the time-sharing approach can significantly reduce packet transmission energy. The reason why a time-sharing  scheme can reduce packet transmission energy is as follows. As we have seen earlier, there is a trade-off between transmission throughput (also the average transmission delay) and packet transmission energy.
If we aim to minimize  packet transmission energy, then the best option is to choose a very low symbol energy level, as illustrated in Figure \ref{fig:fixedcentre_adjustEt}. The price to be paid however is the extremely low throughput. To address this problem, we can send a portion of the packets by using a large symbol energy level to increase the overall throughput. It turns out from our simulation that this time-sharing approach can increase throughput without significantly increasing the packet transmission energy.
%In fact, in some scenarios, we can achieve nearly $65\%$ of energy reduction.

{\bf Remarks: }
The gain from time-sharing depends on a clever choice of $E^{\text{Tx}}_{\, \rm low}$ and $E^{\text{Tx}}_{\, \rm high}$. It also depends on the moving speed of the relay. Usually, the gain is  larger when the relay's moving speed is low. Yet, even when the relay is moving fast, there is still a fairly large reduction in packet transmission energy in the low throughput region.

A different operating characteristic could be achieved by introducing more than two choices for transmission energy level.  Referring to Figures~\ref{fig:mov&jump} and~\ref{fig:timesharen10s1a}, time-sharing allows the system to operate on a curve that connects the two operating points corresponding to
$E^{\text{Tx}}_{\, \rm low}$ and $E^{\text{Tx}}_{\, \rm high}$.  Careful choice of these operating points will ensure the minimum energy time-sharing curve.  More transmission energy levels will result in an operating characteristic with the same or higher total energy.

\begin{figure*}[ht]
\centering
  \includegraphics[scale=0.7]{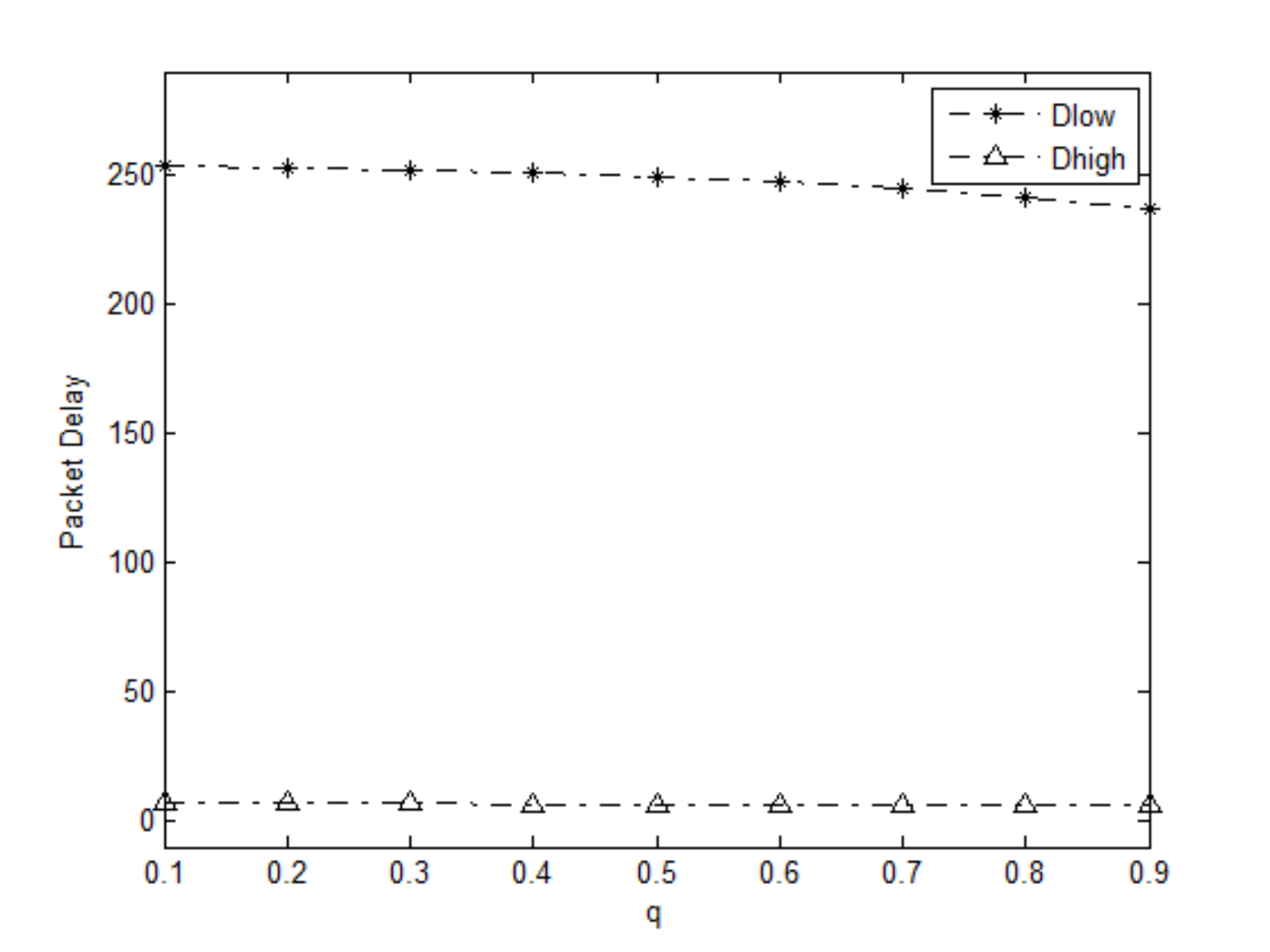}\\
  \caption{Packet delay low energy and high energy in terms of $q$.}\label{fig:line2}
\end{figure*}

If we examine the delay-energy curve (obtained by time-sharing) in Figure \ref{fig:timesharen10s1a}, we can notice that the curve is fairly close to a straight line. In the following, we aim to explain why. Using our Markov chain model, we can easily prove that
\begin{align}
q = \frac{\pi^{*}_{\text{high},0}}{\pi^{*}_{0}}, \text{ and } 1-q  = \frac{\pi^{*}_{\text{low},0}}{\pi^{*}_{0}}
\end{align}
So equation (\ref{eq:totale}) can be rewritten as:
%\begin{eqnarray}
%q &=& \frac{\pi_{\text{high},0}}{\pi_{TS,0}}\\
%1-q & = & \frac{\pi_{\text{low},0}}{\pi_{TS,0}}
%\end{eqnarray}
\begin{eqnarray}
&&  E^{*}_{\text{total}}  \nonumber \\
& &= \left(E^{\text{Tx}}_{\text{low}}  \cdot\frac{ \pi^{*}_{\text{low}}}{\pi^{*}_{\text{low},0}} \cdot \frac{\pi^{*}_{\text{low},0}}{\pi^{*}_{0}} + E^{\text{Tx}}_{\text{high}} \cdot\frac{\pi^{*}_{\text{high}}}{\pi^{*}_{\text{high},0}} \cdot \frac{\pi^{*}_{\text{high},0}}{\pi^{*}_{0}} \right) L
\nonumber \\
& &= \left(E^{\text{Tx}}_{\text{low}} \cdot\frac{\pi^{*}_{\text{low}}}{\pi^{*}_{\text{low},0}} \cdot (1-q) + E^{\text{Tx}}_{\text{high}} \cdot\frac{\pi^{*}_{\text{high}}}{\pi^{*}_{\text{high},0}} \cdot q\right) L \label{eq:Etotalq}
% &=& \biggl(E_{\text{low}} \cdot \tau_{\text{low}} \cdot (1-q) + E_{\text{high}} \cdot \tau_{\text{high}} \cdot q\biggr) L \label{eq:Etotalq}
 \end{eqnarray}
 Let
 \begin{align}
\tau^*_{\text{low}} & =  \frac{\pi^{*}_{\text{low}} }{  \pi^{*}_{\text{low},0}} \\
\tau^*_{\text{high}} & =  \frac{\pi^{*}_{\text{high}} }{  \pi^{*}_{\text{high},0}} 
 \end{align}
Then $\tau^*_{\text{low}}$ (or $\tau^*_{\text{high}}$) is the average delay for a packet to be transmitted using the symbol energy level $E^{\text{Tx}}_{\rm low}$ (or
$E^{\text{Tx}}_{\rm high}$).
Similarly, we can also prove that
\begin{align}\label{eq:13}
\frac{1}{\pi^{*}_{0}} =   \frac{\pi^{*}_{\text{low}}}{\pi^{*}_{\text{low},0}} \cdot (1-q) +  \frac{\pi^{*}_{\text{high}}}{\pi^{*}_{\text{high},0}} \cdot q  .
\end{align}
As the relay is moving, $\tau^*_{\text{low}}$ and $\tau^*_{\text{high}}$ are not constant, but are functions of the time-sharing parameter $q$. In Figure \ref{fig:line2}, we plot the values of $\tau^*_{\text{low}}$ and $\tau^*_{\text{high}}$ in terms of $q$. It turns out that the two values are fairly insensitive to the value of $q$.
Together with \eqref{eq:Etotalq} and  \eqref{eq:13}
this explains why the energy-delay curve in Figure \ref{fig:timesharen10s1a} resembles a straight line.

Before we end this section, we would like to compare our proposed time-sharing approach to a relaying scheme with a sleep mode.
The idea of sleep mode is based on the fact that a node should only be awakened to receive or transmit information when necessary \cite{Ivan2009}. In many data gathering scenarios, it would also be reasonable to turn off the radio and only keep the sensor on for most of the time \cite{kalpakis2003efficient}.
 At a first glance, it seems that the two schemes should perform similarly (in particular when $E^{\text{Tx}}_{\, \rm low}$ is very small in our time-sharing scheme). However, as we shall see,  the two schemes are in fact quite different.

To evaluate the performance of a transmission scheme with a sleep mode, we will specify formally the scheme using our Markov chain model:
Assume that after each successful packet transmission,  there is a probability $p_\text{sleep}$ that the sensor will go to the sleep mode.
Both the sensor $S$ and the relay $R$ will not transmit during the sleep mode. However, the mobile relay will continue to move during the sleep mode.
The duration of when the sensor and the relay will remain in this sleep mode follows the geometric distribution with mean $1/(1-p_\text{sleep}) - 1$.
Once the system is awakened from the sleep mode, it will start transmission of a new packet again.
\begin{figure*}[ht]
\centering
  \includegraphics[scale=0.7]{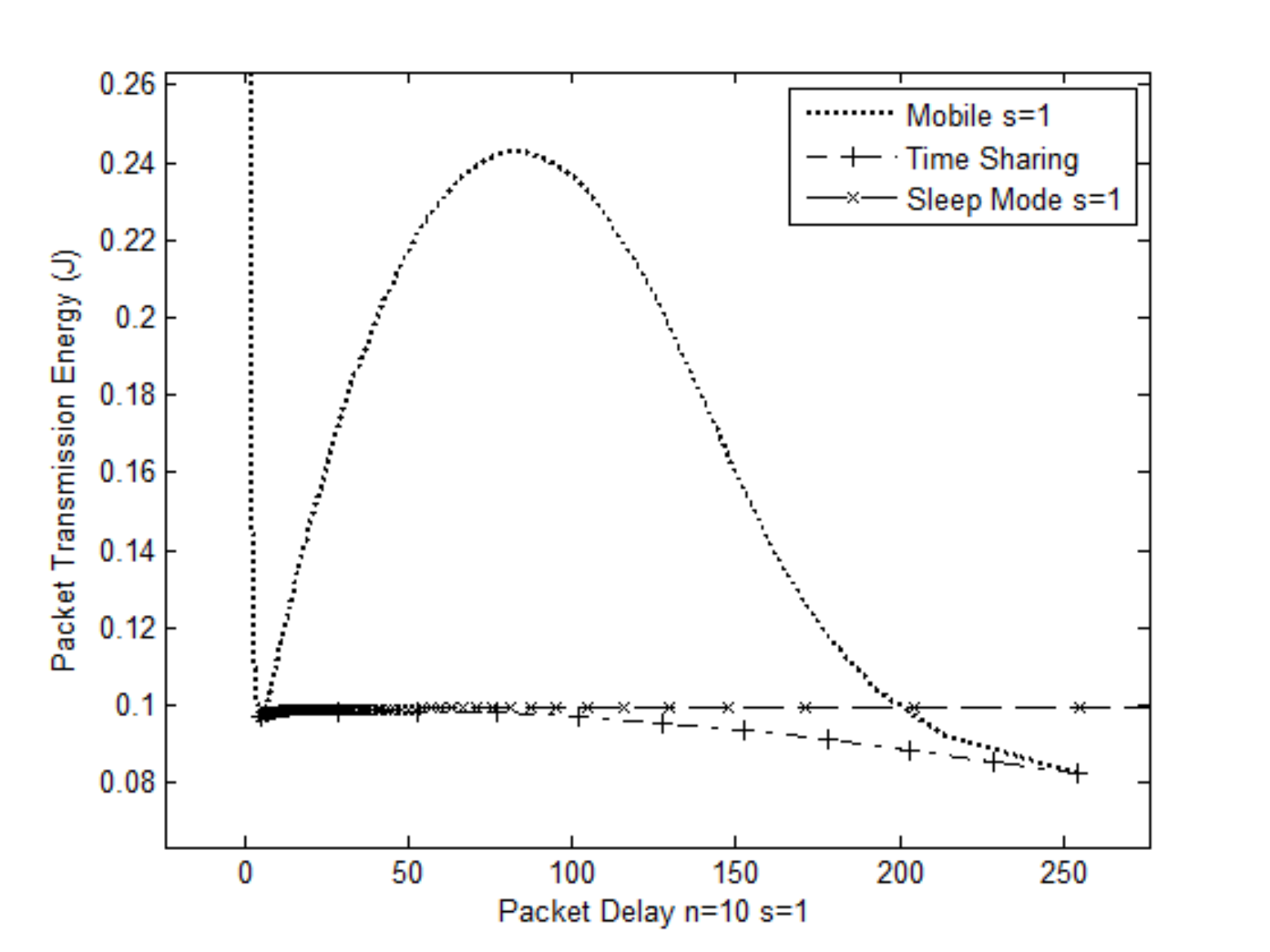}\\
  \caption{Performance comparison of single transmit energy, time-sharing and sleep mode in terms of total energy per packet versus average delay}\label{fig:sleep}
\end{figure*}

If we examine the protocol carefully, we can see right away that a sleep mode enabled scheme cannot reduce packet transmission energy.
It can only lengthen the lifespan of a node by not using all its energy. When the system is awakened and is transmitting a packet, it consumes as much energy as one without a sleep mode.

Figure \ref{fig:sleep} shows the energy-delay curves obtained from our proposed time-sharing scheme and from the sleep mode enabled relaying scheme.
For reference, we also re-plot one of the throughput-energy curves in Figure \ref{fig:mov&jump} as the energy-delay curve.
For our-time sharing scheme, the system parameters are exactly as in Section \ref{sec:char}.
For the scheme with sleep mode, symbol energy $\Et$ is chosen to be $1.8\times10^{-5}$ J, which  is the same as $E^{\text{Tx}}_\text{high}$ in our time-sharing scheme.

As expected, it turns out that transmission scheme with a sleep mode does not help reducing packet transmission energy.
For a transmission scheme with a sleep mode, the packet transmission energy needed is essentially the same as that  without using the sleep mode.

\subsection{Design Principles}
\label{sec:application}
In this section, we summarize the application of lessons learned from our study of the energy and throughput trade-off for two-hop networks with either a fixed or mobile relay. We form a set of operating guidelines or application notes.

We see from our analysis in this paper that having access to a relay with mobility provides us with the opportunity to access the low total energy (very low throughput) operating region for the two-hop network, and furthermore, opens up the possibility of using a time-sharing scheme.
Transmitting with very low power enables the sensor to significantly extend its battery lifetime so long as retransmissions do not dominate the energy consumption. Mobility alleviates the high need for retransmissions at low transmission energies, and allows for successful communication despite a very tight constraint on peak power from the battery.

We found that if we can only adjust the transmission energy level and the mobility of the relay node, the network throughput at low transmissions energies depends a lot on the relay mobility. If the transmission energy is fixed, we can only improve the throughput by increasing the speed of the mobile relay. If the mobility of the relay is fixed, we can select the proper transmission energy to fulfil the requirement of the network throughput.

If the relay mobility is fixed under the time-sharing scheme, we can adjust the selection of time-share energy levels to ensure the delay of the network. If the relay mobility is not fixed, we need to find the optimal time-sharing energy levels to make sure the time scheme is energy efficient in terms of the delay and energy trade-off.

The time-sharing scheme also offers the opportunity to offer different levels of service (i.e.\ priority or preferential treatment) for different packets the network. For example, in many applications, communication of a changed condition is of far greater priority than communication that the status is unchanged.  A time-sharing scheme can easily accommodate the two priority levels.
Sleep mode can be used to extend the lifetime of the network. However, it does not achieve the energy savings per packet that the time-sharing scheme can achieve.

%In many practical application scenarios where users are more interested in observing pre-specified data, we can apply the idea of transmitting different observation by using different transmission energy to save energy.
%Let's say gathered data are usually accepted within a known error bound $[-\varepsilon,+\varepsilon]$,$\varepsilon \in \textbf{R}$.
%
%If the sensor node can predict or estimate the next readings results, and transmits an update to the destination only if the current measurement differs from the predicted data by more than the error bound. The transmit energy consumption will be reduced by avoiding the unnecessary communication.
%
%Obviously, the possible energy consumption saving will depend on the particular sensed phenomenon, the data sampling rate and the used prediction model.

\section{Summary and Future Direction}
\label{sec:concl}
%In this paper we have modeled  a simple a random walk motion based one relay wireless network based on a Markov Chain. We formulated an analysis to minimize the overall energy consumption per packet and network delay subject to other system constraints. The results indicate that the maximum system throughput and minimum system energy consumption
%per packet cannot be achieved simultaneously. There are three different operating regions representing different system throughput and energy consumption trade-offs.
%

In this paper, we have modeled a simple two-hop wireless network with a mobile relay node where transmission energy can be adjusted to improve the performance of the whole network.  We introduced a bounded random walk model for the mobile relay node and characterized the energy and throughput performance using steady-state analysis of the Markov chain model we formulated. Based on observations, we proposed a time-sharing scheme for the low throughput operating region, extending our formulation and analysis for this scenario. We demonstrated the effectiveness of time-sharing in significantly reducing the sensor node energy consumption while achieving the same average throughput and packet delay as a single transmit energy system. The results indicate that time-sharing is an attractive operating mode for the two-hop network with mobile relay.  While the models presented in this paper are simple, the analysis approach developed is broadly applicable.  We have highlighted and discussed possible extensions and generalizations.  Nevertheless, even the simple models were demonstrated to be useful in deriving design principles and operating guidelines for energy conservation in a wireless sensor network.

%More specifically, the time-sharing network scheme should be used for applications which can tolerate significant delay. Since when the sensor node wants to transmit data by using low level transmission energy the mobile relay might be out of the coverage area of S node, large delays may result. If the mobile relay is moving very fast, there will high probability of the relay node to be within the coverage area of both S and D nodes.

%The disadvantage of this network is the increased latency, because both sensor nodes and destinations have to wait for the relays to come close to them. However, for many wireless sensor network applications some latency is acceptable. %We will consider latency, throughput, energy consumption, infrastructure cost in our future model.

\section*{Acknowledgement}
The authors would like to thank the reviewers and the editor for their comments and suggestions, which greatly  help  improving the quality of the paper.

\end{document}